\documentclass[12pt]{article}
\usepackage{xr}

\usepackage{amsmath}
\usepackage{natbib}
\usepackage{url} 

\usepackage{amssymb}
\usepackage{color}
\usepackage{bm}
\usepackage{hyperref}
\usepackage{graphics,graphicx}
\usepackage{algorithmic,algorithm}

\makeatletter

\newcommand{\para}{\bigskip\noindent}

\newtheorem{theorem}{Theorem}[section]

\newtheorem{lemma}[theorem]{Lemma}

\newcommand{\bbeta}{\bm{\beta}}

\newcommand{\blambda}{\bm{\lambda}}

\newcommand{\boldeta}{\bm{\eta}}
\newcommand{\bY}{\mathbf{Y}}

\newcommand{\bX}{\mathbf{X}}

\newcommand{\be}{\begin{equation}}
\newcommand{\ee}{\end{equation}}

\newcommand{\expit}{\text{expit}}

\newcommand{\bs}[1]{\boldsymbol{#1}}

\newcommand{\ML}{\text{ML}}
\newcommand{\Lik}{\text{L}}

\newcommand{\blind}{0}

\begin{document}

\def\spacingset#1{\renewcommand{\baselinestretch}%
{#1}\small\normalsize} \spacingset{1}


%
%
%

\if0\blind
{
  \title{\bf Fast cross-validation for multi-penalty high-dimensional ridge regression}
  \author{Mark A. van de Wiel,  Mirrelijn M. van Nee\thanks{
    The authors acknowledge \textit{NWO-ZonMw TOP grant COMPUTE CANCER (40-00812-98-16012)}}\hspace{.2cm}\\
    Department of Epidemiology and Data Science, Amsterdam University\\ Medical Centers, Amsterdam, The Netherlands\\
    and \\
    Armin Rauschenberger\\
    Luxembourg Centre for Systems Biomedicine (LCSB),\\ University of Luxembourg, Esch-sur-Alzette, Luxembourg}
  \maketitle
} \fi

\if1\blind
{
  \bigskip
  \bigskip
  \bigskip
  \begin{center}
    {\LARGE\bf Title}
\end{center}
  \medskip
} \fi

\bigskip
\begin{abstract}
High-dimensional prediction with multiple data types needs to account for potentially strong differences in predictive signal. Ridge regression is a simple model for high-dimensional data that has challenged the predictive performance of many more complex models and learners, and that allows inclusion of data type specific penalties. The largest challenge for multi-penalty ridge is to optimize these penalties efficiently in a cross-validation (CV) setting, in particular for GLM and Cox ridge regression, which require an additional estimation loop by iterative weighted least squares (IWLS). Our main contribution is a computationally very efficient formula for the multi-penalty, sample-weighted hat-matrix, as used in the IWLS algorithm. As a result, nearly all computations are in low-dimensional space, rendering a speed-up of several orders of magnitude.

We developed a flexible framework that facilitates multiple types of response, unpenalized covariates, several performance criteria and repeated CV. Extensions to paired and preferential data types are included and illustrated on several cancer genomics survival prediction problems. Moreover, we present similar computational shortcuts for maximum marginal likelihood and Bayesian probit regression.
The corresponding \texttt{R}-package, \texttt{multiridge}, serves as a versatile standalone tool, but also as a fast benchmark for other more complex models and multi-view learners.
\end{abstract}

\noindent%
\vfill

\newpage
\spacingset{1.5} 

\section{Introduction}
Many researchers face the challenge of integrating multiple (high-dimensional) data types into one predictive model. For example, in many clinical studies multiple genomics types, such as mutations, gene expression and DNA copy number, are measured for the same individuals. A common task then is to build a model for diagnosing disease or predicting therapy response (both binary outcomes), or for predicting survival. The various data types may differ strongly in predictive signal, which demands a multi-view approach to prediction models. We propose a versatile, computationally efficient framework based on multi-penalty ridge regression.

 Ridge regression \cite[]{Hoerl1970} is one of the oldest statistical models to deal with high-dimensionality. Unlike the more recent sparse methods, it does not select features although posterior selection can be applied \cite[]{Bondell2012, perrakis2019scalable}. Moreover, others have argued that many diseases are likely poly- or even omnigenic \cite[]{boyle2017expanded}, which in combination with high collinearity, may render sparse methods sub-optimal in terms of prediction. The relative good predictive performance of ridge regression compared to several other methods, including its sparse counterpart, lasso, is confirmed in \cite[]{bernau2014cross} for several genomics applications.
On the other side of the spectrum, very dense multi-view deep learners are data hungry in terms of sample size, conflicting with the small $n$ setting in many clinical genomics studies (often $n< 100$ or several hundreds at most), and do not necessarily outperform much simpler models for relatively large genomics data sets either \cite[see][$n \approx 8,000$]{warnat2020scalable}.
Moreover, unlike for many more complicated models, the analytical tractability of ridge regression allows theoretical results on predictive risks \cite[]{dobriban2018high} and model diagnostics \cite[]{ozkale2018logistic}. Finally, the predictive performance of ridge regression can be boosted by incorporating different penalties for groups of features \cite[]{WielGRridge, velten2018adaptive, perrakis2019scalable}.

Our approach aligns with the latter works: it allows different penalties for different data types. The estimation of these, however, is fully based on cross-validation (CV) instead of (empirical) Bayes, rendering the method very generic in terms of i) the response and corresponding model (linear, logistic, Cox); ii) potential inclusion of non-penalized clinical covariates; and iii) evaluation score, such as log-likelihood, area-under-the-roc-curve, and c-index. The versatility of CV is also emphasized by \cite[]{arlot2010survey}, which provides a detailed overview of CV underpinned by theoretical results, such as model consistency.

CV for multiple penalty parameters is computationally demanding, so our aim is to overcome this hurdle. Therefore, we developed \texttt{multiridge}: a method for efficient estimation of multiple penalties by fully exploiting the algebraic properties of the (iterative) ridge estimator. We first shortly review fast CV for ordinary ridge with one penalty parameter \cite[]{hastie2004efficient}, which is useful for initializing the penalties. Then, we combine several matrix identities with a formulation of Iterative Weighted Least Squares (IWLS; used for GLM and Cox ridge regression) in terms of the linear predictors to allow an implementation with only one simple calculation in the high-dimensional feature space per data type: the product of the data matrix with its transpose. This calculation does not need to be repeated for folds of the CV, for different values of the penalties (proposed by the optimizer), or different sample weights in the IWLS algorithm. This renders \texttt{multiridge} very efficient for high-dimensional applications.

To broaden the application scope of \texttt{multiridge} we provide several extensions, including preferential ridge, which accounts for a preference for one data type over the other, and paired ridge, which shrinks parameters of paired covariates (e.g. the same gene measured on both DNA and mRNA). Moreover, we show how to use the presented computational shortcuts to find the penalty parameters with maximum marginal likelihood (MML), instead of CV. The two are compared for several data sets.  Other applications of the computational shortcuts are discussed as well, including kernel ridge regression and Bayesian probit ridge regression fit by a variational algorithm.
We supply details on the implementation, in particular on the optimization strategy for the penalties.
The computational gain of \texttt{multiridge} compared to more naive implementations is illustrated to be several orders of magnitude, which makes it useful as a standalone
tool, but also as a fast benchmark for more complex multi-view models and (possibly sparse) learners. Several variations of \texttt{multiridge} are illustrated on four multi-omics tumor data sets with survival response.

\section{Contributions}
Multi-penalty ridge is not new, and in fact already proposed by \cite{Hoerl1970}. Our aim is to make the model practically usable for modern applications with multiple (very) high-dimensional data types, and linear, binary or survival response. We summarize the main contributions of this paper.

\begin{enumerate}
\item A computationally very efficient formula for the multi-penalty, sample-weighted hat-matrix, crucial in the IWLS algorithm for fitting GLM and Cox ridge regression
\item Extension to inclusion of non-penalized covariates, essential in many clinical high-dimensional prediction problems to account for patient characteristics
\item Efficient CV which avoids repetitive computations in the high-dimensional space
\item Extension of the equality by \cite{perrakis2019scalable}, here \eqref{gamma}, to paired ridge \eqref{gammapaired}
\item Extension to estimation of penalties by marginal likelihood instead of CV
\item \texttt{R}-software \texttt{multiridge} enabling multi-penalty ridge prediction and performance evaluation for a broad array of high-dimensional applications
\item Extension to multi-penalty (prior precision) Bayesian probit ridge regression
\end{enumerate}

\section{Fast uni-penalty CV for ordinary ridge}
Efficient cross-validation for ridge with a single penalty parameter $\lambda$ is discussed by \cite{hastie2004efficient}. Their solution was preceded by \cite{hawkins2002faster}. The latter solution,  however, is memory inefficient, because it requires storing a $p \times p$ matrix. \cite{turlach2006even} offer an efficient alternative based on QR decomposition. This solution, however, is only discussed in the context of linear ridge regression.

Let $\bY$ be the linear response vector of length $n$ (sample size), and $X$ the $n \times p$ design matrix with $p$: the number of covariates.  \cite{hastie2004efficient} show for the ridge estimator that:
\be\label{svd}
\hat{\bbeta}_{\lambda} = (X^TX + \lambda I) ^{-1}X^T \bY = V(R^TR + \lambda I)^{-1}R^T \bY,
\ee where $R$ and $V$ are obtained by singular value decomposition (SVD) of $X$: $X = RV^T = UDV^T$.  This hugely reduces computing time,
as the second estimator requires inverting an $n \times n$ matrix instead of a $p \times p$ one. Importantly, they show that also for CV as well as for GLM and Cox ridge regression only one SVD is required. The last equality in \eqref{svd} breaks down, however, when
$\lambda I$ is replaced by a diagonal matrix $\Lambda$ with non-identical diagonal elements, as required for multi-penalty ridge. If one needs to evaluate only one value of $\Lambda$, this may be solved as in \cite{WielGRridge} by an SVD on $X_{\Lambda} = X\Lambda^{-1/2}$.
Repeating the SVD for many values of $\Lambda$ is computationally costly, however, so therefore we develop an alternative approach.
Nevertheless, fast CV for ordinary ridge is useful to initialize multi-penalty ridge: simply cross-validate each data type separately,
and use the obtained $\lambda$'s as starting values. Supplementary Table \ref{ctsimple} shows that the SVD-based CV implementation can be orders of magnitude faster than plain CV, in particular for large $p$. In all cases, the two methods rendered the exact same penalty.
In case of leave-one-out CV the approximate method introduced by \cite{Meijer2013} can be competitive, in particular for large $n$ and moderate $p$ settings. However, when both $n$ and $p$ are large, SVD-based CV using a lower fold (e.g. 10) seems to be the only reasonable option.

\section{Fast CV for multi-penalty ridge}
Let $X_b$ be the $n \times p_b$ design matrix representing the $b$th data type, $b=1, \ldots, B$. Then, the overall $n \times p, p=\sum_b p_b,$ design matrix $X$ is defined as:
$$X = [X_1 | X_2 | \cdots | X_B].$$
Each $X_b$ corresponds to a penalty $\lambda_b$. The optimal value of $\blambda= (\lambda_b)_{b=1}^B$ will be determined by CV.
Then, denote the penalty matrix by $\Lambda = \text{diag}(\lambda_1 \mathbf{1}_{p_1}, \lambda_2 \mathbf{1}_{p_2}, \ldots, \lambda_B \mathbf{1}_{p_B}),$ with $\mathbf{1}_{p_b}$: a row vector of $p_b$ 1's.
The penalized log-likelihood equals
\begin{equation}\label{penlik}
\ell^{\Lambda}(\bbeta; \bY,X) = \ell(\bbeta; \bY,X) - \frac{1}{2}\bbeta^T \Lambda \bbeta = \ell(\bbeta; \bY,X) - \frac{1}{2}\sum_{b=1}^B \lambda_b ||\bbeta_b||_2^2,
\end{equation}
where the factor 1/2 is used for convenience. This formulation is a special case of generalized ridge regression, already discussed by \cite{Hoerl1970}. \cite{firinguetti1999generalized} suggests an estimator for $\Lambda$, but this only applies to low-dimensional linear models, as it relies on an initial OLS estimator. Our focus lies on efficient estimation of $\Lambda$ by cross-validation in a generic, high-dimensional prediction setting.
Therefore, our aim is to efficiently maximize the cross-validated likelihood \cite[CVL;][]{Houwelingen2006}:
\begin{equation}\label{CVL}
\hat{\Lambda} = \text{argmax}_{\Lambda} \text{CVL}(\Lambda; \bY,X),
\end{equation}
where
$$\text{CVL}(\Lambda; \bY,X) = \sum_{i=1}^n \ell(\hat{\bbeta}^\Lambda_{(-i)}; Y_i,\bX_{i.}),$$
with $$\hat{\bbeta}^\Lambda_{(-i)} = \text{argmax}_{\bbeta}\, \ell^{\Lambda}(\bbeta; \bY_{-f(i)},X_{-f(i)}),$$
and $-f(i)$ denoting removal of all samples in the same fold as sample $i$. Later, we discuss extensions to other performance criteria than CVL.
By definition, the CVL is fully determined by the linear predictor $\bX_i\hat{\bbeta}^\Lambda_{(-i)}$ in GLM and Cox, which will be exploited throughout this manuscript.
All estimates depend on $\Lambda$, so the number of evaluations of $\Lambda$ can be huge:
$n_\Lambda \cdot n_{\text{fold}} \cdot \bar{n}_W ,$ where $n_\Lambda$ is the number of penalty parameter configurations that will be considered by the optimizer, $n_{\text{fold}}$ is the number of folds, and $\bar{n}_W$ is the average number of weight vectors the IWLS algorithm below requires to converge across all folds and $\Lambda$ settings. Hence, each estimate needs to be computed very efficiently.

We fit a ridge-penalized GLM using an iterative weighted least squares (IWLS) algorithm. Cox ridge regression is very similar in spirit (see \cite{Houwelingen2006}; further details are supplied in the Supplementary Material).
Let $\boldeta$ be the linear predictor which we wish to update after initializing it by, for example, $\boldeta = \bf{0}_n$.
Algorithm \ref{IWLS} formulates IWLS in terms of $\boldeta$ in the context of logistic ridge regression.

\begin{algorithm}
Initialize $\boldeta$.\\
Cycle:
\begin{alignat}{2}
\tilde{Y}_i &\leftarrow \expit(\eta_i),  \tilde{\bY} = (\tilde{Y}_i)_{i=1}^n &\text{ (vectorized predictions)}\label{tildeY}\\
w_i &\leftarrow \tilde{Y}_i (1-\tilde{Y}_i), W = \text{diag}((w_i)_{i=1}^{n}) &\text{ (sample weights)}\label{Wdef}\\
\bm{C} &\leftarrow \bY - \tilde{\bY} &\text{ (centered response vector)}\label{Cdef}\\
H_{\Lambda, W} &\leftarrow X (\Lambda + X^T W X)^{-1} X^T  &\text{ (hat matrix)}\label{Hmat}\\
\boldeta_{\Lambda, W} &\leftarrow  H_{\Lambda,W}(\bm{C} + W \boldeta), \boldeta = \boldeta_{\Lambda, W} &\text{ (linear predictor update)}\label{eta}
\end{alignat}
\caption{IWLS algorithm}\label{IWLS}
\end{algorithm}
The updating for $\boldeta$ follows from applying Newton's method (see Supplementary Material).
For other GLMs, one simply needs to replace $\tilde{Y}_i$ and $w_i$ by $E(Y_i|\eta_i)$ and $V(Y_i|\eta_i)$, respectively. Note that one may account for an offset by including it as $\tilde{\bY} = \expit(\boldeta_0 + \boldeta)$. Moreover, note that for linear response $\bY$, no updating and weights are required, as we simply have $\boldeta = H_{\Lambda,I_n}\bY$.

IWLS relies on iterative updating of these quantities until convergence.
Note that the computation of $\boldeta$ does \emph{not} require the large parameter estimate vector $\hat{\bbeta}_{\Lambda}$, but instead relies crucially on the
$n \times n$ hat matrix $H_{\Lambda, W}$. Below we show how to efficiently compute this. When we use CV to find $\Lambda$, many versions of
$H_{\Lambda, W}$ need to be computed, because besides $\Lambda$ and $W$, also the design matrix $X$ varies due to CV. The derivations below show that all these instances of
$(W, \Lambda, X)$ together require to compute the $B$ block-wise sample correlation matrices $S_b = X_b X_b^T$ \emph{only once}. Then, this is the only computation in dimension of order $p$; all other computations are in dimension $n$ or lower. We start with the setting with all $\lambda_b > 0$, which we then extend to allow for an unpenalized block, $\lambda_0 = 0$. Finally, we show how to adapt the calculations in settings where the design matrix $X$ changes, such as for CV and for new samples.

\subsection{All covariates penalized}
Here, assume that all covariates are penalized, so $\forall b: \lambda_b > 0$. We set out to efficiently compute $H_{\Lambda, W} = X (\Lambda + X^T W X)^{-1} X^T $, for possibly many different values of $W$ and $\Lambda$.
We first apply Woodbury's identity to convert matrix inversion of the large $p \times p$ matrix to that of an $n \times n$ matrix plus some matrix multiplications:
 \begin{equation}\label{woodbury1}
 (\Lambda + X^T W X)^{-1} = \Lambda^{-1} - \Lambda^{-1}X^T (W^{-1} + X\Lambda^{-1}X^T)^{-1} X\Lambda^{-1}.
 \end{equation}
The most costly operation in \eqref{woodbury1} is the matrix multiplication $X\Lambda^{-1}X^T$, as $X$ has dimensions $n \times p$. As noted by \cite{perrakis2019scalable}, we have
\begin{equation}\label{gamma}
\Gamma_{\Lambda} =  X\Lambda^{-1}X^T = \sum_{b=1}^B \lambda^{-1}_b \Sigma_b, \text{ with } \Sigma_b = X_b X_b^T.
\end{equation}
This is computationally very useful, because it means that once the $B$ $n \times n$ matrices $\Sigma_b$ are computed and stored,
$\Gamma_{\Lambda} = X\Lambda^{-1}X^T $ is quickly computed for any value of $\Lambda$.
Finally, we have:
\begin{equation}\label{Hmateff}
\begin{split}
H_{\Lambda, W} &= X(\Lambda + X^T W  X)^{-1} X^T\\
&= X \biggl(\Lambda^{-1} - \Lambda^{-1}X^T (W^{-1} + X\Lambda^{-1}X^T)^{-1} X\Lambda^{-1}\biggr) X^T\\
&= \Gamma_{\Lambda} - \Gamma_{\Lambda}(W^{-1} + \Gamma_{\Lambda})^{-1}\Gamma_{\Lambda},
\end{split}
\end{equation}
in which all matrices are of dimension $n \times n$, and hence the matrix operations are generally fast.
The new linear predictor $\boldeta = \boldeta_{\Lambda, W}$ then equals $H_{\Lambda, W} (C + W \boldeta)$, after which \eqref{tildeY} to \eqref{eta} are updated accordingly in the IWLS algorithm.

\subsection{Including unpenalized covariates}
Often it is desirable to include an intercept and also a few covariates without a penalty. E.g. in clinical genomics studies, patient information on age, gender, disease stage, etc. may be highly relevant for the prediction.
In this setting, we need to augment $\Lambda$ with zeros to obtain the entire penalty matrix $\Lambda'$. Then, \eqref{gamma} and \eqref{Hmateff} do not apply, because $\Lambda'^{-1}$ is not defined when at least one of the diagonal elements equals 0. Replacing those by a small penalty may render the matrix calculations instable. Hence, we extended \eqref{Hmateff} to the setting which allows zeros on the diagonal of $\Lambda'$.
First, define:
\begin{equation*}
    X\in\mathbb{R}^{n\times p},\ X_1\in\mathbb{R}^{n\times p_1},\ X_2\in\mathbb{R}^{n\times p_2},\ s.t.\ X=\left[X_1 | X_2\right],
\end{equation*}
and such that $X_1$ contains the covariates left unpenalized and $X_2$ the covariates to be penalized. Here, $X_2$ may consist of multiple blocks
corresponding to penalty matrix $\Lambda$. Then, write the penalty matrix $\Lambda'$, which is rank deficient, as the two-by-two block matrix containing blocks of all zeros and a $p_2\times p_2$ penalty matrix of full rank:
\begin{align}
    \Lambda'= \left[\begin{array}{cc}
    \Lambda'_{11} & \Lambda'_{12}\\
    \Lambda'_{21} & \Lambda'_{22}
    \end{array}\right] = \left[\begin{array}{cc}
    \bs{0}_{p_1\times p_1} & \bs{0}_{p_1\times p_2}\\
    \bs{0}_{p_2\times p_1} & \Lambda
    \end{array}\right].
\end{align}
Furthermore, assume that $X_1$ has linearly independent columns (in the algebraic sense), i.e. $\text{rank}(X_1)=p_1\leq n$. Moreover, let $X_{k,W}= W^{1/2} X_k,$ for $k=1,2,$ and define the projector $P_{1,W} =I_{n\times n} - X_{1,W}(X_{1,W}^TX_{1,W})^{-1}X_{1,W}^T$.
Write \eqref{Hmat} as $H_{W,\Lambda} =  X (X^T W X+\Lambda')^{-1}X^T  = W^{-1/2} X_W (X_W^T X_W+\Lambda')^{-1}X_W^T W^{-1/2}.$

\para
{\bf Proposition 1}\\
\begin{equation}\label{Hmat12}
    H_{W,\Lambda} = W^{-1/2} X_{1,W}(X_{1,W}^TX_{1,W})^{-1}X_{1,W}^T \left( I_{n\times n} - W^{1/2}H_{2,W,\Lambda}W^{1/2}\right) W^{-1/2} + H_{2,W,\Lambda},
\end{equation}
with
\begin{equation}\label{Hmat2}
    H_{2,W,\Lambda} = W^{-1/2} \Gamma_{W, \Lambda}\biggl(I_{n\times n} - (I_{n\times n} + P_{1,W} \Gamma_{W, \Lambda})^{-1}P_{1,W}\Gamma_{W, \Lambda} \biggr)P_{1,W}W^{-1/2},
\end{equation}
where
\begin{equation}\label{gamma2}
\Gamma_{W, \Lambda} =  W^{1/2} \left(\sum_{b=1}^B \lambda^{-1}_b \Sigma_{2,b}\right) W^{1/2}
\end{equation}
and $\Sigma_{2,b}$ as in \eqref{gamma}, with $X_b$ replaced by $X_{2,b}$. The result is proven in the Supplementary Material. Once $\Sigma_{2,b}$ is known for blocks $b= 1, \ldots, B$, all operations in \eqref{Hmat12}, \eqref{Hmat2} and \eqref{gamma2} are on matrices with dimensions $n$ or smaller, facilitating fast computations.

\subsection{Prediction on new samples and final coefficients}
The Supplementary Material shows how to use the results above to perform predictions on new samples, and generate estimated coefficients $\hat{\bbeta}_{\Lambda}$.
The latter does inevitably imply one more multiplication of a $p \times n$ and $n \times n$ matrix, but this needs to be executed only once for
the optimal $\hat{\Lambda}$ and converged weights $W$.

\subsection{Efficient CV}
Use of the Woodbury identity, as in \eqref{woodbury1}, replaces matrices of the form $X^T X$ by $X X^T$. This has a secondary convenient consequence, apart from the much smaller matrices to invert: it allows for very efficient cross-validation.
Computation of the CVL \eqref{CVL} requires the linear predictors for left-out samples. For that \eqref{Hmat} and \eqref{eta} may be used after replacing the first matrix $X$ in \eqref{Hmat} by $X_{\text{out}} = X[\text{out},]$, and the others by $X_{\text{in}} = X[\text{in},]$, where
`out' (`in') denotes the sample indices of the left-out (left-in) samples and $[\text{out},]$ (or $[\text{in},]$) selects the corresponding rows. Then, analogously to \eqref{Hmateff}, we have
\begin{equation}\label{Hmatout}
H_{W,\Lambda,\text{out}} = X_{\text{out}} (\Lambda + X_{\text{in}}^T W X_{\text{in}})^{-1} X_{\text{in}}^T =  \Gamma_{\Lambda,\text{out},\text{in}} - \Gamma_{\Lambda,\text{out},\text{in}}(W^{-1} + \Gamma_{\Lambda,\text{in},\text{in}})^{-1}\Gamma_{\Lambda,\text{in},\text{in}},
\end{equation}
with $\Gamma_{\Lambda,\text{out},\text{in}} = X_{\text{out}}\Lambda^{-1}X_{\text{in}}^T = \sum_{b=1}^B \lambda_b^{-1}\Sigma_{b,\text{out},\text{in}}$ and
$\Sigma_{b,\text{out},\text{in}} = X_b[\text{out},] (X_b[\text{in},])^T$. $\Gamma_{\Lambda,\text{in},\text{in}}$ and $\Sigma_{b,\text{in},\text{in}}$ are defined analagously. These are also required within the IWLS algorithm, as this is now applied to the training (`in') samples.
These matrices are conveniently obtained from $\Sigma_{b} = X_b X_b^T$, which is computed only once from the entire data set, because:
$$\Sigma_{b,\text{out},\text{in}} = \Sigma_{b}[\text{out},\text{in}], \quad \text \quad \Sigma_{b,\text{in},\text{in}} = \Sigma_{b}[\text{in},\text{in}],$$
and therefore also:
\begin{equation}\label{gammasub}
\Gamma_{\Lambda,\text{out},\text{in}} = \Gamma_{\Lambda}[\text{out},\text{in}], \quad \text \quad \Gamma_{\Lambda,\text{in},\text{in}} = \Gamma_{\Lambda}[\text{in},\text{in}].
\end{equation}
Likewise, this applies to the setting with an unpenalized data block for the computation of the cross-validation versions of $H_{2,W,\Lambda}$ \eqref{Hmat2}
and $\Gamma_{W, \Lambda}$ \eqref{gamma2}.

\subsection{Complexity}
Let us consider the reduction in complexity due to the presented computational shortcuts. First, \eqref{woodbury1} reduces the complexity of matrix inversion (by standard algorithms) from $\mathcal{O}(p^3)$ to $\mathcal{O}(n^3)$. Then, the remaining most complex operation (assuming $p>n$) is the matrix multiplication $X\Lambda^{-1}X^T$, with
complexity $\mathcal{O}(pn^2)$ when computed directly. As this is linear in $p$, this equals the total complexity of computing $X\Lambda^{-1}X^T$ by using $\Sigma_b = X_b X_b^T$, for $b=1, \ldots, B$, as in \eqref{gamma}. The huge advantage of the latter, however, is that it requires only one such calculation, as outlined above. Naive recalculation of
$X\Lambda^{-1}X^T$ requires repeating it for all configurations of weights $W$ and \text{folds} (both change $X$) and $\Lambda$ (which changes during optimization).
The number of such configurations visited during IWLS, CV and $\Lambda$-optmization easily runs into several thousands. Our algorithm does require storing $B$ $n \times n$ matrices $\Sigma_b$.
For most practical applications this is not a limiting factor, however, as the number of data types $B$ is usually limited.

\section{Computing times for simulated cases}
\begin{figure}
\includegraphics[scale=0.45]{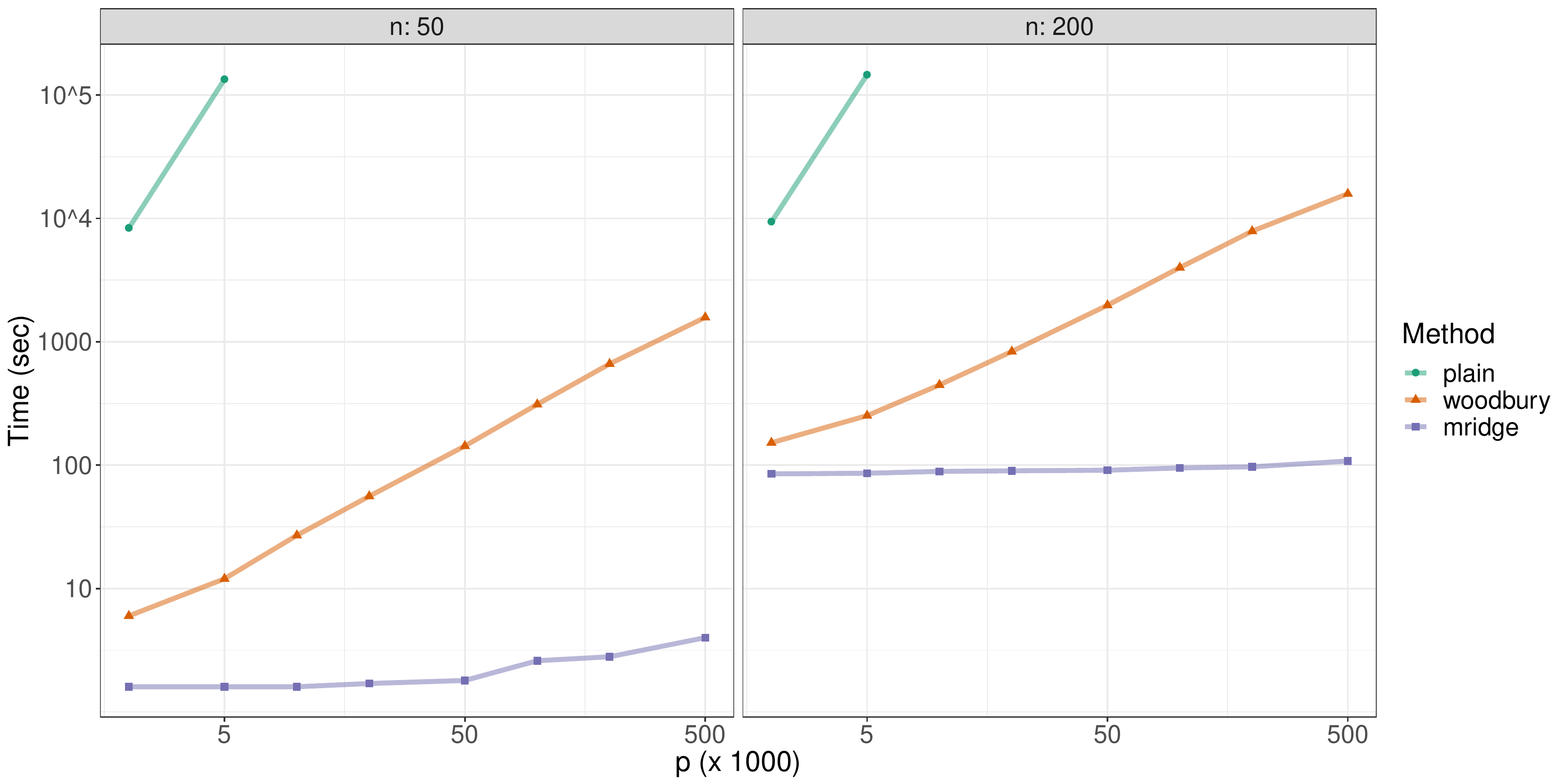}
\caption{Computing times for 1,000 evaluations of $(\Lambda, W, \text{Folds})$}\label{fig:comp}
\end{figure}

Figure \ref{fig:comp} displays computing times on a PC with a 2.20GHz core for several simulated cases. We show results for $n=50$ and $n=200$, and varying $p$, where
$p = 2 + 2*p_{\text{pen}}$, with 2: the number of unpenalized covariates and $p_{\text{pen}}$: the size of each of the $B=2$ penalized data blocks.
We assume 1,000 evaluations of $(\Lambda, W, \text{folds})$.  E.g. for $K$-fold CV with $K=10$ and for an average number of IWLS iterations of 5, this implies evaluation of 1000/(5*10) = 20 different values of $\Lambda$ by the optimizer. This is still fairly modest, and for $B \geq 2$ often a larger number of $\Lambda$'s will be evaluated.  The figure clearly shows the substantial computational benefit of \texttt{multiridge} w.r.t. to plain ridge fitting and ridge fitting that uses Woodbury's identity only.


\section{Implemented extensions}
Below we present several extensions available in the software \texttt{multiridge}. Some of these will be illustrated for the data applications.
\subsection{Extension 1: Repeated CV}
To increase stability, some have argued for \emph{repeated} CV to optimize the penalty parameters \cite[]{boulesteix2017ipf}. A theoretical argument for repetition of subsampling is found in \cite{fongmarginal}, who establish an equivalence with marginal likelihood optimization. Our software allows repeated CV. Computing time is linear in the number of repeats, so the computational shortcuts presented here are important for practical use of repeated CV.


\subsection{Extension 2: Alternative CV utility criteria}
We generally recommend to use cross-validated (partial) likelihood as the utility criterion for maximizing $\Lambda = \text{diag}({\blambda})$, as this is consistent with the likelihood criterion used to fit the model, given $\Lambda$.
Moreover, it is a smooth function, which benefits optimization algorithms, unlike ranking-based alternatives like area-under-the-roc-curve (AUC).  Nevertheless, in some cases one may want to explicitly maximize an alternative
criterion, e.g. because it is better interpretable than CVL, or to match it with the predictive performance criterion used to evaluate and compare the predictors with alternative ones.
Our software allows the user to supply a user-specific utility function, to be combined with any of the optimizers available in \texttt{R}'s \texttt{optim} function. Generally, for non-smooth utilities like AUC, optimizers require more evaluations of the objective function than for smooth ones, and may converge to a local optimum. The latter problem can be alleviated by repeated CV, although the resulting profile is likely flatter than that of CVL (see Supplementary Figure \ref{fig:profile}), which may cause the optimizer to terminate too early. In any case, the larger number of evaluations required by alternative, non-smooth utility criteria warrants the usefulness of the computational shortcuts presented here.

\subsection{Extension 3: Paired ridge}
For some applications, the covariates in $X$ are paired, and one may want to make use of this information by coupling the corresponding parameters.
A well-known example are genes measured on mRNA and DNA for the same samples, in which case one may want to couple parameters corresponding to the same gene. Another example is two transformations or representations of the same data, e.g. continuous gene expression and binarized gene expression, low versus high.
We assume paired columns are juxtaposed in $X$: $X= (\bX_{1,.1}, \bX_{2,.1}, \bX_{1,.2}, \bX_{2,.2}, \ldots,\bX_{1,.p}, \bX_{2,.p})$, where $\bX_{k,.j}$ denotes the $j$th column
of design matrix $X_k, k=1,2$.
Then, in penalized likelihood \eqref{penlik} paired parameters $(\beta_j, \beta'_j)$ correspond to $(\bX_{1,.j},\bX_{2,.j})$ and $\Lambda$ is now a block diagonal matrix:

$$\Lambda = I_p \otimes \Lambda_s =  I_p \otimes \left[\begin{array}{cc}
    \lambda_{1} & -\lambda_{3}\\
    -\lambda_{3} & \lambda_{2}
    \end{array}\right],
    $$
because one wishes to employ a paired penalty:
\be\label{pairedpen}\tilde{\lambda}_1\beta_j^2 + \tilde{\lambda}_2(\beta'_j)^2 + \tilde{\lambda}_c (\beta_j-\beta'_j)^2 =
\lambda_1\beta_j^2 + \lambda_2(\beta'_j)^2 - \lambda_3 \beta_j\beta'_j,
\ee with $\lambda_1 = \tilde{\lambda}_1+\tilde{\lambda}_c, \lambda_2 = \tilde{\lambda}_2 + \tilde{\lambda}_c, \lambda_3=\tilde{\lambda}_c$.
An alternative formulation for the paired penalty is $\tilde{\lambda}_c (\tilde{\lambda}_1^{1/2}\beta_j-\tilde{\lambda}_2^{1/2}\beta'_j)^2,$ which scales the $\beta$'s with their prior standard deviations, i.e.
$1/\tilde{\lambda}_k^{1/2}, k=1,2$, rendering the pairing possibly more natural. The two formulations, however, are equivalent, as
one obtains the right-hand side of \eqref{pairedpen} by setting $\lambda_1 = \tilde{\lambda}_1(1+\tilde{\lambda}_c), \lambda_2 = \tilde{\lambda}_2(1 + \tilde{\lambda}_c), \lambda_3=(\tilde{\lambda}_1\tilde{\lambda}_2)^{1/2}\tilde{\lambda}_c$. Note that the latter equivalence is useful for initialization: suppose that $\tilde{\lambda}_k$ has been determined by fast uni-penalty CV, and one initializes $0 < \tilde{\lambda}_c < 1$ as the relative paired penalty (e.g. $\tilde{\lambda}_c= 1/4$), then the implied initial values of $(\lambda_1, \lambda_2, \lambda_c)$ should be roughly on the correct scale.

In this paired setting, we conveniently have $\Lambda^{-1} = I_p \otimes \Lambda_s^{-1} =: \Omega$.
Write $$\Omega_s = \Lambda_s^{-1} = \left[\begin{array}{cc}
    \omega_{1} & \omega_{3}\\
    \omega_{3} & \omega_{2}
    \end{array}\right].$$
Then,
\be\label{gammapaired}
\Gamma_\Lambda = X\Lambda^{-1} X^T = X\Omega X^T = \omega_1 X_1 X_1^T + \omega_2 X_2 X_2^T + \omega_3 X Q X^T,\quad Q = I_p \otimes \left[\begin{array}{cc}
    0 & 1\\
    1 & 0
    \end{array}\right].
\ee
Therefore, also in this setting computations are very efficient once the $n \times n$ matrices $\Sigma_k = X_k X_k^T, k=1,2$ and $\Sigma_Q = X Q X^T$
are computed and stored. Here, $Q$ is a large $2p \times 2p$ matrix, but does not need to be generated for computing $\Sigma_Q$: $Q$ is a permutation matrix, so $X Q$ is simply computed by swapping the paired columns in $X$.
These results trivially extend to triplets or larger blocks of covariates. As long as these blocks are small, inversion of $\Lambda$ is fast and so is the computation of $\Gamma_{\Lambda}$.
It also extends easily to combinations with block(s) of unpaired covariates: simply apply \eqref{gammapaired} to the paired covariates and add this to \eqref{gamma}, which is applied to the unpaired covariates.

With a small modification, the above also applies to settings where $X_1$ and $X_2$ are measured on \emph{different} individuals, and one still wishes to couple paired parameters.
For example, when two studies have a similar set-up and measure the same, or very similar covariates, possibly on slightly different platforms. Then, create $X$ by juxtaposing columns $\bm{X}^0_{1,.j} = (X_{1,1j}, \ldots, X_{1,n_1j}, \mathbf{0}_{n_2})^T$ and $\bm{X}^0_{2,.j} = (\mathbf{0}_{n_1},X_{2,1j}, \ldots, X_{2,n_2j})^T,$ so 0's are inserted for individuals for which either $X_1$ or $X_2$ does not apply. Then, the two covariate sets correspond to two different parameter estimates, but these are shrunken towards one another.

Finally, note that fusing is another mechanism to couple parameters, e.g. when an ordering is known, such as DNA position of a gene \cite[]{Chaturvedi2014fused}. This leads to a banded $\Lambda$ matrix. Efficient algorithms exist to compute the inverse of such matrices, which, however, is generally dense. The latter precludes efficient (repeated) computation of $\Gamma_{\Lambda} =  X\Lambda^{-1}X^T$.

\subsection{Extension 4: Preferential ridge}
Sometimes, one or more particular data types may be preferable over others, in particular when one believes that these data types generalize better to other settings. A well-known example in genomics is the preference for relatively stable DNA-based markers (copy number, mutations, methylation) over mRNA-based ones (gene expression). In a linear elastic net setting, \cite{aben2016tandem} propose a two-stage approach, \texttt{TANDEM}:
first, an elastic net with the prefered markers is fitted; second, an elastic net with the non-prefered (gene expression) markers is fit to the residuals of the first model. The authors show that such a strategy can perform competitively to the baseline strategy without preferential markers, while arguing that the two-stage model is likely more robust. The competitive performance was also observed by \cite{klau2018priority}, who employ a similar approach using a multi-stage lasso.

One could follow the same two-stage regression fitting in a ridge setting. The main strength of ridge regression, however, is that it deals well with collinearity among covariates. Such collinearity may be rather strong within as well as \textit{across} data types (e.g. DNA copy number and mRNA expression of the same gene are often strongly correlated in tumor profiles). This would be ignored when fitting the regression in two stages. Therefore, we prefer to fit one ridge regression in the end, but reflect the preference by first estimating $\blambda$ using only the preferential data types, $\mathcal{P} \subset \{1,..,B\}$. Then, the corresponding penalties, $\blambda_{\mathcal{P}} = (\lambda_b)_{b \in \mathcal{P}}$, are fixed, and $\blambda_{\mathcal{P}^{C}}= (\lambda_b)_{b \notin \mathcal{P}}$ is optimized conditional on $\blambda_{\mathcal{P}}$ using all covariates in the regression.
We present a data example of this strategy below.


\section{Software notes}
\subsection{Implementation, data and scripts}
The method is implemented in \texttt{R}-package \texttt{multiridge}, available at \url{https://github.com/markvdwiel/multiridge}. This also links to a demo \texttt{R}-script and to data and scripts used to produce the results in this manuscript. The demo includes an example on predicting survival from multi-omics data for patients suffering from mesothelioma cancer, which is extensively discussed below. In addition, it includes a classification example on diagnostics of a pre-stage of cervical cancer using $p_1 = 699$ miRNAs and $p_2 \approx 365,620$ methylation markers, shortly discussed further on. This demonstrates the computational efficiency of \texttt{multiridge} for large $p$. Further details and references on the data are supplied in the Supplementary Material. To accommodate performance assessment on data, \texttt{multiridge} contains a function to execute double CV, which adds another loop of computations to the problem.

\subsection{Checks}
Results for estimating $\bbeta$ (and hence $\boldeta = X\bbeta$) have been checked for correctness against the plain (but computationally inefficient) generalized ridge estimator for the linear regression case,  $\hat{\bbeta}_{\Lambda} = (X^TX + \Lambda)^{-1} X^T \bY$.
These checks are also available as part of the code.


\subsection{Optimization}
We use \texttt{R}'s  \texttt{optim} function to optimize CVL \eqref{CVL} w.r.t. $\blambda$, the unique diagonal elements of $\Lambda$. The aforementioned SVD-based uni-penalty CV is used to initialize all $B$ components of $\blambda$.
There is no guarantee that the optimization problem is convex in terms of $\blambda$. Therefore, our default optimization strategy after initialization is to first search globally by running a simulated annealer (\texttt{SANN}) using a limited number of iterations, e.g. 10. Then, we apply a more local method, defaulting to \texttt{Nelder-Mead} for multi-penalty  optimization and \texttt{Brent} for uni-penalty optimization. In our experience, this strategy is robust against potential local optima, while still computationally efficient.

\subsection{Parallel computing}
The algorithm is parallelized at the level of the folds by distributing the computation of the contribution of each fold to the cross-validated likelihood \eqref{CVL} for given $\Lambda$. Note that this parallelization is very efficient, because \eqref{gammasub} implies that the largest object to be distributed across nodes is the $n \times n$ matrix $\Gamma_{\Lambda}$.

\section{Application: survival prediction with TCGA data}

We use \texttt{multiridge} to predict overall survival from multi-omics data stored in The Cancer Genome Atlas (TCGA): microRNA expression (miRNA; $p_1 \approx 1,400$), messenger RNA gene expression (mRNA; $p_2 \approx 19,000$), and DNA copy number variation (CNV; $p_3 \approx 10,000$).  Four tumor types were studied: Mesothelioma (MESO; $n=84$), Kidney renal clear cell carcinoma (KIRC; $n=506$), Sarcoma (SARC; $n=253$) and Thymoma (THYM; $n=118$).
These four were chosen for their variability in sample size (but all larger than $n>80$), availability of matched miRNA, mRNA and CNV data, and for showing some signal. Other (large) sets such as breast cancer and ovarian cancer (BRAC and OV) rendered no or a weak signal at best (c-index $< 0.6$) for any of the methods below, and hence these results were not shown.
The data were preprocessed as described in \cite{rauschenberger2019sparse}. Further details on the data and the Cox ridge model are given in the Supplementary Material.

The purpose of these analyses is three-fold: i) to report computing times for \texttt{multiridge} on real data with variable sample sizes; ii) to compare standard \texttt{multiridge} with preferential ridge (\texttt{multiridge\_pref}), and with elastic net based benchmarks; and iii) to illustrate paired ridge (\texttt{multiridge\_pair}), and compare it with \texttt{multiridge}. In all studies, an outer 3-fold CV-loop was used to evaluate performance on test samples. The training-test splits were balanced in terms of events (deaths). The training involved an 10-fold CV-loop for optimizing the penalty parameters.

\subsection{Application of preferential ridge and its benchmarks}
For preferential ridge, we used miRNA and CNV as prefered data types and mRNA gene expression as secondary one. The first two are generally prefered in clinical practice for their better stability and robustness. As benchmarks we applied two variations of \texttt{TANDEM} \cite[]{aben2016tandem}, which is only implemented for linear regression. \texttt{TANDEM} uses a two-stage elastic net (EN), where the second stage fits the non-prefered markers to the residuals of the elastic net fitted to the prefered markers. Equivalently (in the linear setting), we introduce the $n$ linear predictors of the first stage as offsets for the second stage elastic net. Advantage of this approach is that it also applies to GLM and Cox regression, which we need here. The first variation (\texttt{EN2}) uses miRNA and CNV together in the first stage, and mRNA in the second stage. The second variation (\texttt{EN3}) is a three-stage approach, using the order CNV $\rightarrow$ miRNA $\rightarrow$ mRNA, which, unlike \texttt{EN2}, allows different penalties for miRNA and CNV. We use \texttt{glmnet} to fit the elastic nets with defaults as in \texttt{TANDEM} ($\alpha=0.5$ and $\lambda_1 = \text{lambda.1se}$).
A final simple benchmark is the lasso applied to all data types simultaneously. We use \texttt{glmnet} to fit it with default $\lambda_1 = \text{lambda.min}$.

\subsection{Computing times}
Table \ref{ctime} shows computing times
for \texttt{multiridge} and \texttt{multiridge\_pref} using four nodes of a Windows x64 server, build 14393.
Computing times were averaged over three data splits. Variation was small, hence not
shown. We distinguish computing times for initialization (using the efficient SVD-based
CV for each omics-type separately) from those of the two multi-penalty approaches. For the latter
two, the maximum number of optimization iterations was set to 10 and 25 for the global
and local search by \texttt{SANN} and \texttt{Nelder-Mead/Brent}, respectively. We observe that computing
times are very reasonable, even for the largest sample size, $n_\text{train} = 337$. Note that for linear
models computing times are likely substantially shorter as no IWLS iterations are needed; in
addition, we noticed that logistic ridge is often faster than Cox ridge, as it tends to require
fewer IWLS iterations. Finally, as a side note: the elastic net approaches fitted by \texttt{glmnet} (including CV of the penalty) roughly took 1.5-3 times more computing time, without parallel computation however.

\subsection{Comparative performance}
\begin{figure}
\includegraphics[scale=0.45]{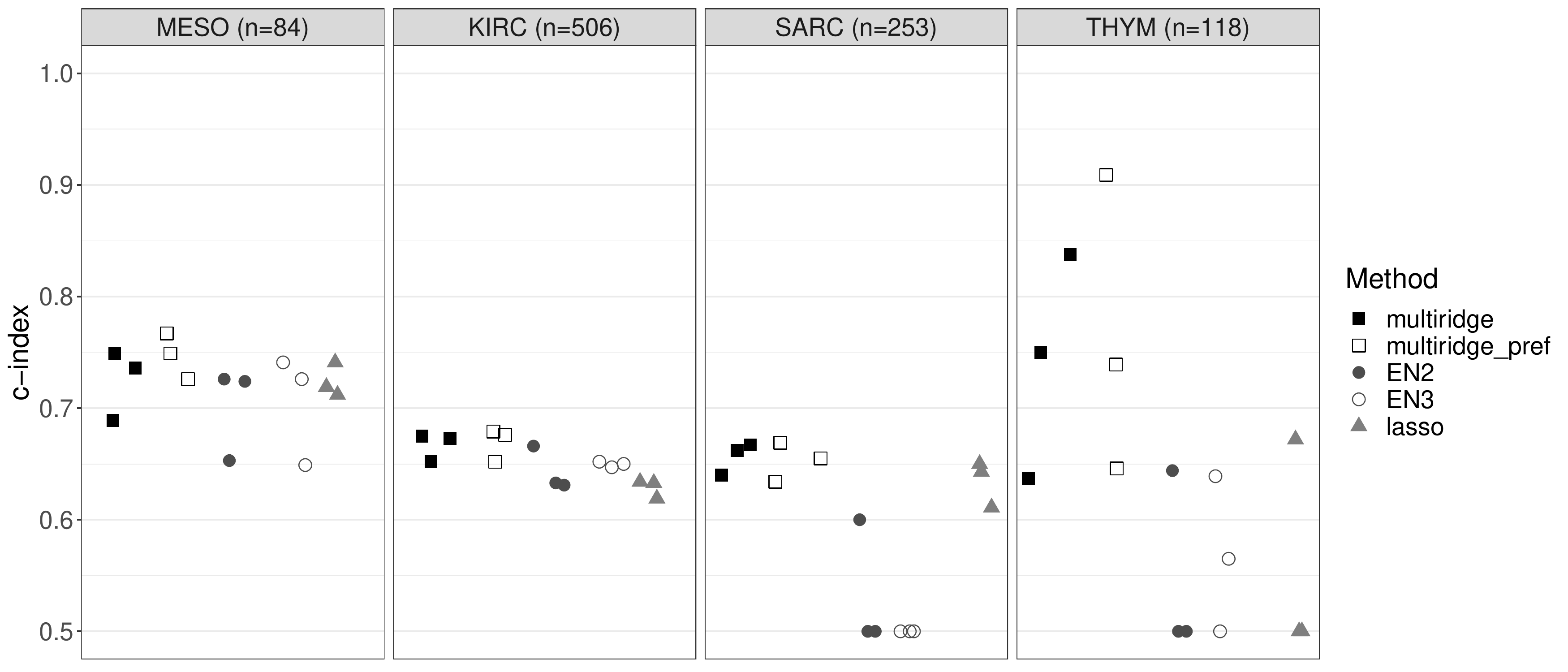}
\caption{Predictive performance for three splits per data set, as measured by c-index}\label{fig:compare}
\end{figure}

Figure \ref{fig:compare} shows the comparative performances of \texttt{multiridge, multiridge\_pref, EN2, EN3, lasso}, as evaluated
on the three test sets by the c-index. Alternatively, we used log-likelihood as evaluation criterion. This rendered
qualitatively similar results, so these are not shown. First, we corroborate the results in
\cite[]{aben2016tandem} that the preferential approach (here, \texttt{multiridge\_pref}) can indeed render very
competitive results to the standard, non-preferential approach (\texttt{multiridge}). Second, we observe
markedly better performance for the ridge-based methods than for the EN models for at least two out of four
tumor types, SARC and THYM. Third, the ridge-based models are also better than \texttt{lasso}
for at least two out of four tumor types, KIRC and THYM.
The EN and lasso models, however, have the advantage of selecting covariates, which may counterweigh the small performance loss for the other two data sets.
For the THYM data set, performances vary substantially between splits due to the small
number of events: 9.
Finally, Table \ref{penalties} shows that preferential ridge (\texttt{multiridge\_pref}) generally penalizes miRNA less
than \texttt{multiridge} does, while penalizing the non-prefered marker, mRNA gene expression, much more. This leads
to substantially larger (smaller) regression coefficients for miRNAs (mRNA) for \texttt{multiridge\_pref}, as
shown in Supplementary Table \ref{absbetas}.

\begin{table}[ht]
\centering
\begin{tabular}{r|rrrr}
  \hline
 & $n_{\text{train}}$ & Init & \texttt{multiridge} & \texttt{multiridge\_pref} \\
  \hline
MESO & 56 & 2.23 & 5.93 & 9.42 \\
  KIRC & 337 & 87.14 & 102.30 & 143.56 \\
  SARC & 168 & 14.65 & 14.81 & 24.02 \\
  THYM & 78 & 2.58 & 7.76 & 15.73 \\
   \hline
\end{tabular}
\caption{Computing times (sec) for initialization, \texttt{multiridge} and \texttt{multiridge\_pref}}\label{ctime}
\end{table}

\begin{table}[ht]
\centering
\begin{tabular}{r|rrr||rr|r}
  \hline
 & \multicolumn{3}{|c||}{\texttt{multiridge}} & \multicolumn{3}{c}{\texttt{multiridge\_pref}}\\
 & miRNA & CNV & mRNA & miRNA  & CNV & mRNA \\
  \hline
  MESO & 2103  & 34068 & 8840 & 394  & 81201 & 15521\\
  KIRC & 1334 & 105079 & 44512 & 834  & 1.7e+6 & 64291\\
  SARC & 4048  & 13036 & 16472 & 2254  & 13398 & 20359 \\
  THYM & 47  & 19376 & 14631 & 73  & 58108 & 2.2e+9 \\
   \hline
\end{tabular}\caption{Median penalties (across three splits) per data set and type}\label{penalties}
\end{table}

\subsection{Application of paired multiridge}
As in ordinary regression settings, the best scale to represent a given data type is not known beforehand. In fact, in an adaptive elastic net setting, the joint use of a continuous and binary representation was shown to be potentially beneficial \cite[]{rauschenberger2019sparse} for omics-based tumor classification. Our default multi-ridge (\texttt{multiridge}) allows for including both representations using different penalties to reflect different predictive signal for the two representations. In addition, paired multi-ridge (\texttt{multiridge\_pair}) specifically accounts for the pairing by the paired penalty \eqref{pairedpen}.

\begin{figure}
\includegraphics[scale=0.45]{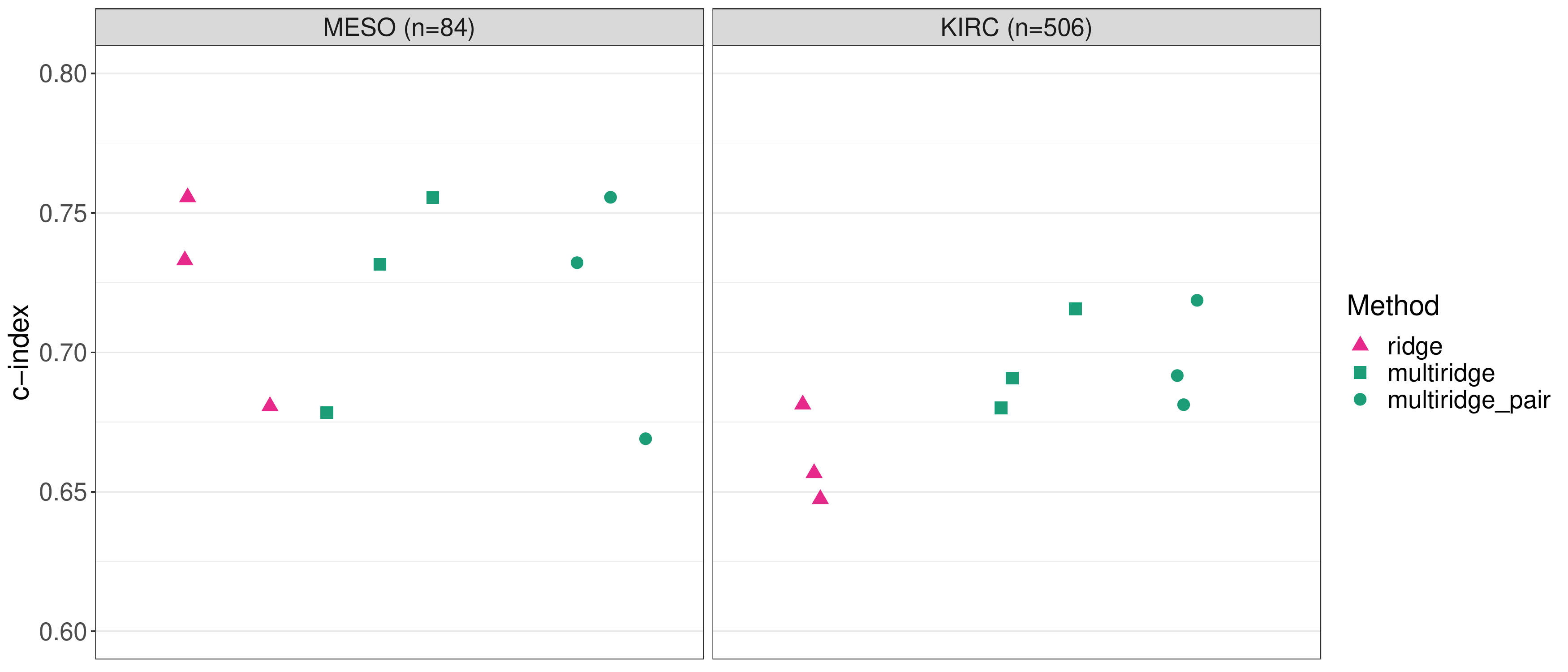}
\caption{Predictive performance for three splits per data set, as measured by c-index}\label{fig:perfpaired}
\end{figure}

\begin{figure}
\includegraphics[scale=0.45]{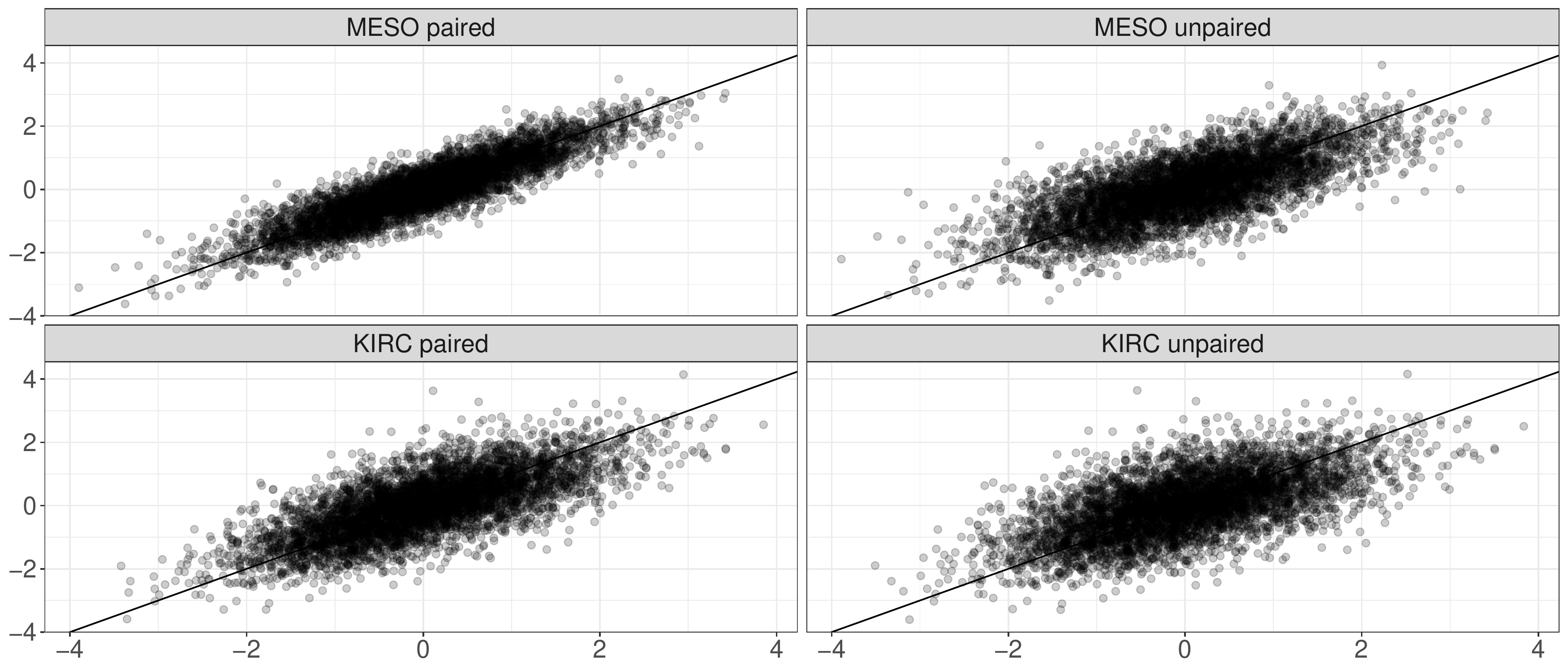}
\caption{Standardized estimated $\beta$'s for the continuous (x-axis) and binarized (y-axis) expression values }\label{fig:betas}
\end{figure}

Using the MESO and KIRC mRNA gene expression data, we assessed whether augmenting this data with its binarized counterpart improved prediction of survival. Per gene, we thresholded the  expression data at the median to render a $(-1,1)$, i.e. low vs high, representation. Less then 10\% of genes showed very little variation on this binary scale (genes with many zero counts), and were therefore filtered out. We noted that computing times were very similar to those in the first two rows of Table \ref{ctime}, and hence these are not further detailed here. Figure \ref{fig:perfpaired} shows the predictive performances for i) ordinary \texttt{ridge} using the continuous gene expression data only; ii) \texttt{multiridge} on both data representations with two unpaired penalties; and iii) \texttt{multiridge\_pair} with an additional paired penalty.
We observe that performances are very much on par for the MESO data, whereas for the large KIRC set adding the binarized representation improves prediction. Pairing does not further improve the predictive performance here. Figure \ref{fig:betas}, however, shows that
\texttt{multiridge\_pair} has an edge in terms of interpretation: the paired penalty clearly increases the agreement between the two parameter estimates
corresponding to the same gene (results shown for first training-test split; results for other splits were very similar).
Correlations between estimates averaged across splits are: 0.72 (0.90) for the MESO data and 0.58 (0.71) for the KIRC data using \texttt{multiridge} (\texttt{multiridge\_pair}), with very limited variability across splits. \cite{Chaturvedi2014fused} reach a similar conclusion for fused lasso, which fuses parameters of neighboring genes on the genome: no or little improvement for prediction, but better interpretability due to increased agreement for estimates of neighboring genes.

\section{Alternative to CV: maximum marginal likelihood}
A prominent alternative to CV is maximum  marginal likelihood (ML). The marginal likelihood is in general not analytical, so approximations are required. \cite{wood2011fast} presents Laplace approximations for the ML for a large variety of models and penalization schemes, implemented in the \texttt{R} package \texttt{mgcv}.
Unfortunately, \texttt{mgcv} currently does not run on high-dimensional data. As mentioned before, a standard solution is to
apply SVD to $X_{\Lambda} = X\Lambda^{-1/2}$. Given $\Lambda$, the maximum of the approximate ML is efficiently found by \texttt{mgcv}. Hence, repeating SVD for the potentially large matrices $X_{\Lambda}$ is the main computational hurdle for this approach.

We developed a second solution, which is computationally superior to the SVD-based solution.  As shown in the Supplementary Material, the marginal likelihood can be written in terms of the $n$-dimensional linear predictor $\boldeta$ with a multivariate Gaussian prior with mean $\mathbf{0}$ and covariance matrix $\Gamma_\Lambda = X \Lambda^{-1} X^T.$ Then, as before, $\Gamma_\Lambda$ is very efficiently computed for many values of $\Lambda$ using \eqref{gamma}. This means that for any $\Lambda$ we may use
\texttt{mgcv} to efficiently compute the ML by simply defining the design matrix to be an $n$-dimensional identity matrix representing the linear predictors, and
a penalty matrix $\Gamma_\Lambda^{-1}$ using the \texttt{paraPen} argument in \texttt{mgcv}'s main function \texttt{gam}. Then, for optimizing $\Lambda$ we use the same strategy as for CV, as described above. We verified that the SVD-based solution and the second solution, referred to as \texttt{multiridge\_ML}, indeed rendered the exact same solutions for $\Lambda$. The latter, however, was 10-15 times faster when applied to the four TCGA multi-omics data sets with survival response. Once the hyper-parameters are estimated with ML we use \texttt{multiridge} with given $\Lambda$ to issue predictions on new samples and to estimate $\bbeta$.

The preferred and paired settings can also be used with ML. This is trivial for the first: it simply implies fixing the penalties once optimized for the preferred data types. The latter, pairing, is accommodated by specifying penalty matrix $\Gamma_\Lambda^{-1}$, resulting from \eqref{gammapaired}. Maximum ML is implemented in the \texttt{multiridge} package, including these two options. Unpenalised variables, however, can not yet be used with ML (see Discussion).

\subsection{Comparison \texttt{multiridge\_ML} with \texttt{multiridge\_CV}}
We compared \texttt{multiridge\_ML} with \texttt{multiridge\_CV} on the basis of three criteria: 1) Computational efficiency; 2) Predictive performance; and 3) Estimation (of $\bbeta$) performance.
\subsubsection{Computing time}
Both methods are equally efficient in the high dimension $p$, as they both use \eqref{gamma}, after which all computations are in dimension $n$.    Then, \texttt{multiridge\_ML} is more efficient in $n$: both fitting algorithms (\texttt{mgcv} and \texttt{IWLS}) use Newton-Raphson for optimization, but \texttt{multiridge\_ML} does not require multiple fits as it does not apply cross-validation. Empirically, we found that for the four TCGA multi-omics data sets with survival response \texttt{multiridge\_ML} was $\approx 3$ times faster than \texttt{multiridge\_CV} with 10-fold CV.

 \subsubsection{Predictive performance}
\begin{figure}
\includegraphics[scale=0.45]{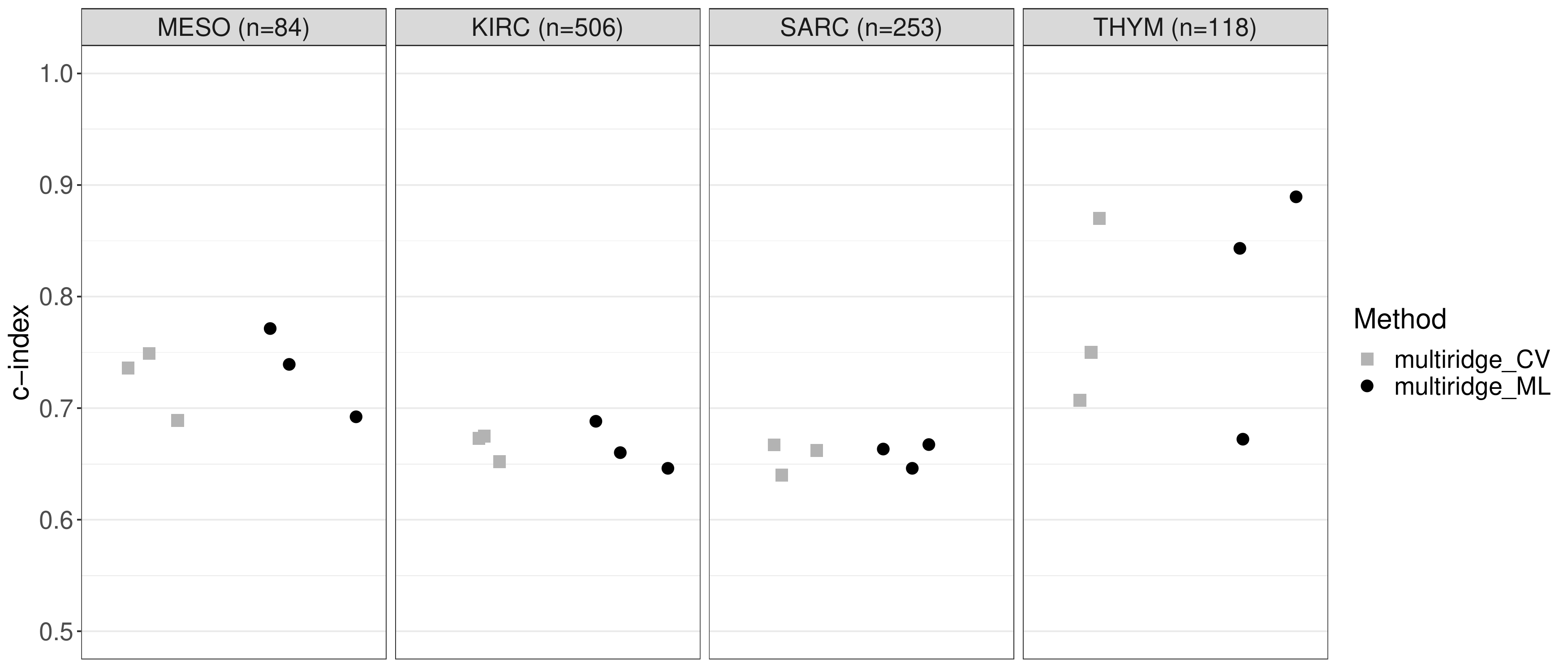}
\caption{Predictive performance for three splits per data set, as measured by c-index}\label{fig:perfmgcv}
\end{figure}

 We applied \texttt{multiridge\_ML} to the four TCGA multi-omics data sets with survival response, and evaluated its performance in exactly the same way as we did for \texttt{multiridge\_CV}. Figure \ref{fig:perfmgcv} compares the predictive performances of \texttt{multiridge\_ML}  with that of \texttt{multiridge\_CV} for four TCGA multi-omics data sets with survival response. It shows that the two are very competitive in terms of predictive performance.

We extend the comparison to logistic regression. As we do not have binary response we generated it using the three omics blocks $X_1, X_2, X_3$ (representing miRNA, mRNA, CNV), and simulated coefficients
 $\bbeta= (\bbeta_{1}, \bbeta_{2},\bbeta_{2})$. Then, we generated each component of $\bbeta_{b}$ i.i.d. from $N(0,1/\lambda_b)$ for $b = 1, 2, 3.$ We used two settings for $\blambda = (\lambda_1,\lambda_2, \lambda_3): \blambda_1 = (20,1000,10000); \blambda_2 = c(100,200,2000)$. Then, $Y_i \sim \text{Bern}(p_i)$, with $p_i = \exp(\eta_i)/\exp(1+\eta_i), \eta_i = \sum_{b=1}^3 \mathbf{X}_{b,i.} \bbeta_b.$ We use the same three training-tests splits as for the survival setting.
 For evaluating predictive performances, we use area-under-the-ROC curve (AUC); log-likelihood on the test sets rendered very similar conclusions. Supplementary Figures 3 and 4 show the AUCs for both simulations. Again, the two methods are very competitive for both settings.

\subsubsection{Estimation performance}
As we know the true $\blambda$ and $\bbeta$ in the simulated logistic setting, we may compare their estimands with the true values. We focus on the latter, as
these allow interpretation on covariate level. Note that in a very high-dimensional, dense setting (with severe collinearities) accurate absolute estimation of $\bbeta$ is a too high ambition. However, accurate relative estimation may be feasible and is very relevant for: a) interpretation of groups of covariates; b) posterior selection procedures \cite[such as][]{Bondell2012}; c) posterior analyses that rely on covariate rankings, such as gene set enrichment \cite[]{subramanian2005gene}.

\begin{table}[ht]
\centering
{\renewcommand{\arraystretch}{0.7}
\begin{tabular}{rrrrr||rrrr}
  \hline
   & ML & CV & ML & CV   & ML & CV & ML & CV\\
   & \multicolumn{2}{c}{500}  & \multicolumn{2}{c||}{1000}  & \multicolumn{2}{c}{500}  & \multicolumn{2}{c}{1000}\\

  \hline
 & \multicolumn{4}{c||}{MESO}  & \multicolumn{4}{c}{MESO}\\
Test1 &  51 &  51 & 145 & 155  &   0 & 182 &   0 & 647\\
  Test2 &  27 &  43 &  81 & 151 & 187 & 192 & 660 & 656 \\
  Test3 &  18 &  44 &  45 &  87 & 196 & 187 & 681 & 674 \\\hline
   & \multicolumn{4}{c||}{KIRC}  & \multicolumn{4}{c}{KIRC} \\
  Test1 &  28 &  41 &  79 & 103  & 178 & 175 & 661 & 657\\
  Test2 &  11 &  36 &  37 &  82  & 161 & 163 & 653 & 664\\
  Test3 &  38 &  14 &  87 &  48  & 174 & 175 & 652 & 647\\\hline
   & \multicolumn{4}{c||}{SARC}  & \multicolumn{4}{c}{SARC}\\
  Test1 &  52 &  54 & 126 & 156  & 164 & 180 & 489 & 669\\
  Test2 &  28 &  54 &  86 & 121  & 181 & 187 & 607 & 648\\
  Test3 &  53 &  54 & 163 & 161  & 180 & 176 & 662 & 661\\\hline
   & \multicolumn{4}{c||}{THYM} & \multicolumn{4}{c}{THYM}\\
  Test1 &  16 &  45 &  76 & 151 &   0 & 163 &   4 & 638\\
  Test2 &  43 &  45 & 116 & 157 & 158 & 156 & 639 & 640\\
  Test3 &  11 &  10 &  44 &  45 & 166 & 170 & 588 & 646\\
   \hline
\end{tabular}
}\caption{Number of overlapping estimated and true $\beta$'s in top 500, 1000 for both ML and CV. Left: $\blambda_1$, right: $\blambda_2$}\label{overlap}
\end{table}
%
%
Therefore, we ranked the true values of $|\beta|$, and the corresponding estimated values.
Table \ref{overlap} shows the overlap for either ML or CV when we `select' the top 500 or top 1000 out of $\approx 30,000$ coefficients in total. We observe that for this criterion
\texttt{multiridge\_CV} is either competitive or superior to \texttt{multiridge\_ML} across almost all data sets and training-test splits.
The cause for the inferior estimation performance of \texttt{multiridge\_ML} may be the very strong overestimation of $\blambda$ with respect to the true value:
in setting 1 the median (1rst quartile, 3rd quartile) overestimation fold across the 4 data sets is 42 (8,172), while it is 16 (5,31) for \texttt{multiridge\_CV}.
In setting 2 these were 26 (9,106) and 12 (4,29), respectively.

\subsubsection{Classification performance for cervical cancer example}
Finally, we also evaluated a classification example on diagnostics of a pre-stage of cervical cancer using $p_1 = 699$ miRNAs and $p_2 = 365,620$ methylation markers,
with $n=43$ samples (25 controls and 18 cases). Details are described in the Supplementary Material. Here, the computational gain is enormous: our method requires
only one multiplication of the large methylation matrix with itself. Therefore, running \texttt{multiridge} took just a few seconds on a PC, whereas naive cross-validation would increase computing time by a factor of several hundreds.
Supplementary Figure 4 displays the ROC-curves for \texttt{multiridge\_CV}, \texttt{multiridge\_ML} and \texttt{lasso}, with respective AUCs: 0.666, 0.474, 0.558.
Hence, \texttt{multiridge\_CV} performs superior to the other two here. Possibly, the Laplace approximation used in \texttt{multiridge\_ML} for the marginal likelihood is less accurate for this binary, small sample setting.

\section{Further extensions}
We outline two extensions of the proposed method that are not part of \texttt{multiridge}.
\subsection{Kernel ridge regression}
Kernel ridge is a well-known extension to ordinary ridge regression in machine learning. In our setting, \eqref{gamma} shows that multi-penalty ridge predictions depend on the data type $b=1, \ldots, B$ only via the scaled $n \times n$ sample covariance matrices $\Sigma_b = X_bX_b^T$.  Our software, \texttt{multiridge}, easily allows for the use of kernels: simply replace $\Sigma_b$ by $\Sigma^K_b = K(X_b, X_b^T),$ where $K$ is a kernel. Here, the inner product is referred to as the linear kernel. A consequence of decomposition \eqref{gamma} is that different kernels may be used for different data types, at no or little additional computational cost. In particular, a linear kernel may not be optimal for potentially unbalanced binary data types (such as mutational status). E.g. when using a conventional $(0,1)$ coding we have $0 \cdot 0 = 0 \cdot 1 = 0$. That is, according to the linear kernel, the 5-feature mutational profile $[0 0 0 0 0]$ is as close to $[1 1 1 1 1]$ as it is to $[0 0 0 0 0]$. Note that while an alternative binary coding like $(-1,1)$ solves this issue, it leads to another one: it would value an agreement between samples on occurrence of the (say rare) event (denoted by a `1', like a mutation) as much as agreement of the non-occurrence (-1). For potentially better agreement measures one may either turn to well-known statistical measures as Cohen's kappa or the Jaccard index. Alternatively, one may try to learn a kernel, preferably in a semi-supervised framework to make use of large numbers of unlabeled observations \cite[]{Salakhutdinov2008}. Note that while non-linear kernels can be used in our framework to produce predictions, they do not (easily) allow interpretation on the covariate level, because
the translation to coefficients ($\bbeta$) is lost. E.g. the popular Gaussian kernel expands to all polynomials terms of the covariates.


\subsection{Bayesian multi-penalty probit ridge regression }
In the Supplementary Material we demonstrate that equality \eqref{gamma} is also useful to derive a very efficient algorithm for obtaining posteriors from Bayesian probit ridge regression. First, we reformulate the variational Bayes (VB) algorithm in \cite[]{ormerod2010explaining} to the $n$-dimensional (hence low-dimensional) space. Next, we show that
this iterative algorithm can also be formulated in terms of updating the hat-matrix $H_{\Lambda, I_n}$, which is efficiently computed during the iterations.
Moreover, we show that the evidence lower-bound (elbo), often used in VB algorithms to monitor convergence or to estimate hyper-parameters, can be expressed in terms of low-dimensional quantities too. These quantities are efficiently computed for multiple $\Lambda$'s using \eqref{gamma} and \eqref{Hmateff}. All-in-all,
this demonstrates the broad applicability of the computational shortcuts presented here.

\section{Discussion}
Multi-penalty ridge is a multi-view penalized method. Several other methods estimate multiple penalty parameters in penalized regression. For sparse settings, \cite{boulesteix2017ipf} introduced \texttt{IPF-lasso}, which cross-validates all penalties in a lasso setting. As one often does not know whether a sparse or dense setting applies, a reasonable strategy in practice may be to apply both methods to the data at hand, and compare the predictive performances. If a small predictive model is desirable, post-hoc variable selection may be applied to the ridge model \cite[]{Bondell2012}.

We also provide marginal likelihood (ML) maximization, the formal empirical Bayes criterion, as a promising alternative to CV, in particular from the computational, perspective. For logistic regression, however, the performance was sometimes inferior to that of CV. This is possibly due to inferior accuracy of Laplace approximation of the ML for logistic regression, because of the very discrete response. Possibly, dedicated adjustments, e.g. by using skew normal distributions, my improve results \cite[]{ferkingstad2015improving}. Finally, the CV-based method is more flexible as it can also optimize non-likelihood criteria, and allows to include non-penalized (fixed) variables. The latter is not straightforward for ML as adding fixed covariates renders the number of covariates larger than the number of samples, which is not allowed by \texttt{mgcv}. A solution for this is left as future work.

Estimation of hyperparameters may alternatively be performed by moment-based (empirical) Bayes methods \cite[\texttt{GRridge}]{WielGRridge}. This method adaptively estimates hyperparameters for possibly many groups of covariates. When  testing \texttt{GRridge}, however, we noticed that moment-based empirical Bayes estimation of the penalties was inferior to CV when applied to groups that represent different data types, in particular  when dimensions differ substantially. The two ridge-based methods should, however, integrate well. Hence, a future research direction is to merge multi-penalty estimation across (\texttt{multiridge}) and within (\texttt{GRridge}) data types.

Multi-view or integrated learners are not necessarily better than learners that use one single data type, in particular when assessed on one's own data. In clincial genomics, it is often difficult to improve RNA-based predictions by adding other genomic data types \cite[]{aben2016tandem}. \cite{broet2009prediction} showed, however, that integrating DNA-based markers with RNA-based ones lead to a more robust classifier with a lower generalization error. This is also a reason to allow a preference for particular data type(s) in \texttt{multiridge}.

We focused on high-dimensional prediction for clinical studies, in which sample sizes are usually small to modest at most. For large $n$ applications,
say several thousands or more, other algorithms than IWLS, such as stochastic gradient descent, may be more efficient. Whether and how this applies to
our setting, including unpenalized covariates, non-linear and possibly censored response and the estimation of multiple penalties, is left as a future research direction.

Our computational shortcuts apply also to Bayesian counterparts of multi-penalty ridge regression. For probit ridge regression we developed an $n$-dimensional version of a variational algorithm by \cite{ormerod2010explaining}, and showed how to efficiently estimate the penalties by expressing the lower bound for the marginal likelihood in terms of the hat-matrix. The results is an efficient Bayesian multi-view classifier for high-dimensional data. It complements \texttt{multiridge}, with the additional benefit of providing uncertainties of parameters and predictions.

We realize that \texttt{multiridge} is based on a simple, one-layer model. Hence, more advanced models may certainly outperform \texttt{multiridge} for data that supports sparse, non-linear, or multi-layer representations. This often comes at the price of reduced interpretation, and sometimes also with (partly) subjective hyperparameter tuning, which may lead to inferior generalization. In any case, a quick comparison with \texttt{multiridge} allows one to judge whether the margin in terms of performance counterbalances the differences in complexity, interpretation, and level of subjectivity.
To conclude, we believe the use of \texttt{multiridge} is two-fold: first, for users as a versatile, interpretable stand-alone multi-view learner; and second, for developers as a fast benchmark for more advanced, possibly sparse, models and multi-view learners.

\section{Supplementary Material}

\subsection{Computing times CV single penalty ridge}
Below we present computing times for plain CV, fast CV using SVD, and approximated leave-one-out CV (LOOCV) as discussed in \citep{Meijer2013}. Table \ref{ctsimple} presents results for LOOCV and 10-fold CV.
Data sets used are: 1) the methylation data set \texttt{dataFarkas} as available in the \texttt{R}-package \texttt{GRridge} \cite[]{WielGRridge}, which has dimensions
$n \times p = 37 \times 40,000$ and binary response; 2) the MESO miRNA data set presented in the main document, with dimensions $84 \times 1,398$ and survival response; 3) as 2) but mRNA, with dimensions $84 \times 19,252$; 4) the KIRC miRNA data set presented in the main document, with dimensions $506 \times 1,487$ and survival response; 5) as 4) but mRNA, with dimensions $506 \times 19,431$.

\begin{table}[ht]
\centering
\begin{tabular}{r||rr||rrr||rr}
  \hline
data & \multicolumn{2}{c||}{Dimensions} & \multicolumn{3}{c||}{LOOCV} & \multicolumn{2}{c}{10-fold}\\
set & $n$ & $p (\times 1,000)$ & Plain & SVD & Approx & Plain & SVD \\
  \hline
1 & 37 & 40.0 &  46.96 & 1.81 & 12.09 & 20.35 & 0.77  \\    \hline
2 & 84 & 1.4 & 26.75 & 13.46 & 3.98 & 4.35 &  1.95 \\    \hline
3 & 84 & 19.3 & 200.08 & 13.27 & 23.12 & 31.04 &  2.22 \\    \hline
4 & 506 & 1.5 & $>$2,000 & $>$2,000 & 85.47 & 154.78  & 112.97 \\    \hline
5 & 506 & 19.4 & $>$2,000 & $>$2,000 & $>$2,000 & 1,713.50 &  146.62 \\    \hline
\end{tabular}\caption{Computing times in seconds }\label{ctsimple}
\end{table}

\subsection{Equivalence IWLS with Newton updating}

The IWLS updating for $\bbeta = (\beta_j)_{j=1, \ldots, p}$ is well-known to be equivalent to Newton updating:

$$\bbeta^{\text{new}} = \bbeta - \mathbf{H}^{-1} \mathbf{g},$$
where $\mathbf{H}$ and $\mathbf{g}$ are the Hessian and gradient function of the objective function (here penalized log-likelihood), respectively.
Therefore,
$$\boldeta^{\text{new}} = \boldeta - X \mathbf{H}^{-1} \mathbf{g}.$$
We have
\begin{equation}\label{derivatives}
\begin{split}
\mathbf{H} &= -X^T W X - \Lambda,\\
\mathbf{g} &=  X^T \bm{C} - \Lambda\bbeta,
\end{split}
\end{equation}
with $\bm{C} = \bY-\tilde{\bY}$, i.e. the response centered around the current prediction.
Substitution renders
\begin{align*}
\boldeta^{\text{new}} &= \boldeta + X ( X^T W X + \Lambda)^{-1}X^T \bm{C} - X ( X^T W X + \Lambda)^{-1}\Lambda\bbeta\\
&= H_{\Lambda,W}\bm{C}  + X(\bbeta - ( X^T W X + \Lambda)^{-1}\Lambda\bbeta)\\
&= H_{\Lambda,W}\bm{C} +  X (X^T W X + \Lambda)^{-1})((X^T W X + \Lambda)\bbeta - \Lambda\bbeta)\\
&= H_{\Lambda,W}\bm{C} +  X (X^T W X + \Lambda)^{-1}X^T W X\bbeta\\
&= H_{\Lambda,W}(\bm{C} + W\boldeta)
\end{align*}
Inclusion of a fixed intercept (e.g. $\boldeta_0 = \text{logit}(\bar{\bY})$) has no impact on the above, as it would not be part of $\mathbf{H}$ and $\mathbf{g}$.

%

\subsection{Proof of Proposition 1: unpenalized variables}
Write \eqref{Hmat} as $H_{W,\Lambda} =  X (X^T W X+\Lambda')^{-1}X^T  = W^{-1/2} X_W (X_W^T X_W+\Lambda')^{-1}X_W^T W^{-1/2}.$
We now derive Proposition 1: an alternative, computationally efficient, expression for $H_{W,\Lambda}$.
For notational convenience we first drop the $W$ index for matrices.  Define:
\begin{equation*}
    X\in\mathbb{R}^{n\times p},\ X_1\in\mathbb{R}^{n\times p_1},\ X_2\in\mathbb{R}^{n\times p_2},\ s.t.\ X=\left[X_1 | X_2\right],
\end{equation*}
and such that $X_1$ contains the covariates left unpenalized and $X_2$ the covariates to be penalized. Therefore, we write the penalty matrix $\Lambda'$, which is rank deficient, as the two-by-two block matrix containing blocks of all zeros and a $p_2\times p_2$ penalty matrix of full rank:
\begin{align}
    \Lambda'= \left[\begin{array}{cc}
    \Lambda_{11} & \Lambda_{12}\\
    \Lambda_{21} & \Lambda_{22}
    \end{array}\right] = \left[\begin{array}{cc}
    \bs{0}_{p_1\times p_1} & \bs{0}_{p_1\times p_2}\\
    \bs{0}_{p_2\times p_1} & \Lambda.
    \end{array}\right]
\end{align}
Furthermore, assume that $X_1$ has linearly independent columns, i.e. $\mathrm{rank}(X_1)=p_1\leq n$ (which is reasonable as one would not include two predictors that are perfectly collinear as unpenalized covariates).

\subsection{Goal} We are interested to represent the following matrix, $L\in\mathbb{R}^{p\times n}$ consisting of the matrices $L_1\in\mathbb{R}^{p_1\times n}$ and $L_2\in\mathbb{R}^{p_2\times n}$, in low-dimensional space $n$:
\begin{align}
    L:= \left[\begin{array}{c}
    L_1 \\
    L_2
    \end{array}\right] :=(X^TX+\Lambda')^{-1}X^T = \left[\begin{array}{cc}
    X_1^TX_1 & X_1^TX_2\\
    X_2^TX_1 & X_2^TX_2+\Lambda
    \end{array}\right]^{-1}\left[\begin{array}{c}
    X_1^T \\
    X_2^T
    \end{array}\right]
\end{align}

\subsection{Result} Define:
$$ P_1 =I_{n\times n} - X_1(X_1^TX_1)^{-1}X_1^T,\\
$$ which is the orthogonal projector onto the kernel (or null space) of $X_1^T$, or equivalently on the orthogonal complement of the column space of $X_1$.
Then, $L_1,L_2$ are given by:
\begin{align}\label{L1L2}
    L_1 &= (X_1^TX_1)^{-1}X_1^T \left( I_{n\times n} - X_2L_2\right)\\
    L_2 &=\left(\Lambda^{-1} - \Lambda^{-1}X_2^T(I_{n\times n} + P_1 X_2 \Lambda^{-1} X_2^T)^{-1} P_1 X_2 \Lambda^{-1}\right) X^T_2P_1.
\end{align}

\subsection{Derivation}
In deriving the expressions, we use two lemmas given below; the two by two block matrix inversion lemma and Woodbury's inversion lemma.
\begin{lemma} (two by two block matrix inversion)
\begin{align*}
    \left[\begin{array}{cc}
    A & B \\
    C & D
    \end{array}\right]^{-1} = \left[\begin{array}{cc}
    A^{-1} + A^{-1}B(D-CA^{-1}B)^{-1}CA^{-1} & -A^{-1}B(D-CA^{-1}B)^{-1} \\
    -(D-CA^{-1}B)^{-1}CA^{-1} & (D-CA^{-1}B)^{-1}
    \end{array}\right]
\end{align*}
\end{lemma}

\begin{lemma} (Woodbury's matrix inversion, for singular $C$)
\begin{equation*}
    (A + BCD)^{-1}=A^{-1}-A^{-1}B(I + CDA^{-1}B)^{-1}CDA^{-1}
\end{equation*}
\end{lemma}

Use the two by two block matrix inversion lemma with
\begin{align*}
    A:=X_1^TX_1,\ B:=X_1^TX_2,\ C:=B^T,\ D:=X_2^TX_2+\Lambda
\end{align*}
Define $P_1$ as above; $P_1=I_{n\times n} - X_1(X_1^TX_1)^{-1}X_1^T$. Then we find:
\begin{align*}
    L_1 &= (A^{-1} + A^{-1}B(D-CA^{-1}B)^{-1}CA^{-1})X_1^T + (-A^{-1}B(D-CA^{-1}B)^{-1})X_2^T\\
    &= \left[A^{-1} + A^{-1}X_1^TX_2\left(X_2^TX_2+\Lambda -X_2^TX_1A^{-1}X_1^TX_2\right)^{-1}X_2^TX_1A^{-1}\right]X_1^T \\
    &\qquad - \left[A^{-1}X_1^TX_2(X_2^TX_2+\Lambda-X_2^TX_1A^{-1}X_1^TX_2)^{-1}\right]X_2^T\\
    &= A^{-1}X_1^T + A^{-1}X_1^TX_2(X_2^TX_2+\Lambda-X_2^TX_1A^{-1}X_1^TX_2)^{-1}X_2^T \\
    &\qquad \cdot \left(X_1A^{-1}X_1^T - I_{n\times n}\right)\\
    &= A^{-1}X_1^T \left[ I_{n\times n} + X_2(\Lambda+X_2^T\left(I_{n\times n} -X_1A^{-1}X_1^T\right)X_2)^{-1}X_2^T\right.\\
    &\qquad \cdot \left.\left(X_1A^{-1}X_1^T - I_{n\times n}\right)\right]\\
    &= A^{-1}X_1^T \left( I_{n\times n} - X_2(\Lambda+X_2^TP_1X_2)^{-1}X_2^T P_1\right)\\
    &= (X_1^TX_1)^{-1}X_1^T \left( I_{n\times n} - X_2L_2\right)\\
    L_2 &= -(D-CA^{-1}B)^{-1}CA^{-1}X_1^T + (D-CA^{-1}B)^{-1}X_2^T \\
    &= -(D-CA^{-1}B)^{-1}X_2^TX_1(X_1^TX_1)^{-1}X_1^T + (D-CA^{-1}B)^{-1}X_2^T \\
    &= (D-CA^{-1}B)^{-1}X_2^T\left(I_{n\times n} -X_1(X_1^TX_1)^{-1})X_1^T \right)  \\
    &= \left(X_2^TX_2+\Lambda-X_2^TX_1(X_1^TX_1)^{-1}X_1^TX_2\right)^{-1}X_2^T\left(I_{n\times n} -X_1(X_1^TX_1)^{-1})X_1^T \right)  \\
    &= \left(X_2^T\left(I_{n\times n} - X_1(X_1^TX_1)^{-1}X_1^T\right)X_2+\Lambda\right)^{-1} X_2^T \left(I_{n\times n} -X_1(X_1^TX_1)^{-1})X_1^T \right)  \\
    &= \left(X_2^TP_1X_2+\Lambda\right)^{-1} X_2^T P_1 \\
    &= \left(\Lambda^{-1} - \Lambda^{-1}X_2^T(I_{n\times n} + P_1 X_2 \Lambda^{-1} X_2^T)^{-1} P_1 X_2 \Lambda^{-1}\right) X^T_2P_1,
\end{align*}
where the last equality follows from Woodbury's matrix inversion lemma.

Now, we apply the results for $L1, L2$ to matrices $X_{1,W} = W^{1/2} X_1$ and $X_{2,W} = W^{1/2} X_2$, and denote results for these matrices by subscript $W$. First, let
\begin{equation}
\Gamma_{W, \Lambda} = X_{2,W}\Lambda^{-1}X_{2,W}^T = W^{1/2} X_2 \Lambda^{-1}X_2^T W^{1/2} = W^{1/2} \left(\sum_{b=1}^B \lambda^{-1}_b \Sigma_{2,b}\right) W^{1/2},
\end{equation}
with $\Sigma_{2,b} = X_{2,b}^T X_{2,b}.$

Then,
\begin{equation}\label{Hmatunpen}
\begin{split}
    H_{W,\Lambda} &=   W^{-1/2} X_W (X_W^T X_W+\Lambda')^{-1}X_W^T W^{-1/2}\\
    &= H_{1,W,\Lambda} + H_{2,W,\Lambda}\\
    &= W^{-1/2} X_{1,W}L_{1,W,\Lambda}W^{-1/2} + W^{-1/2} X_{2,W} L_{2,W,\Lambda}W^{-1/2}\\
    &=  W^{-1/2} X_{1,W}(X_{1,W}^TX_{1,W})^{-1}X_{1,W}^T \left( I_{n\times n} - X_{2,W}L_{2,W,\Lambda}\right)W^{-1/2} \\
    &\qquad + W^{-1/2} X_{2,W} L_{2,W,\Lambda}W^{-1/2}\\
    &= W^{-1/2} X_{1,W}(X_{1,W}^TX_{1,W})^{-1}X_{1,W}^T ( I_{n\times n} -  W^{1/2}H_{2,W,\Lambda}W^{1/2})W^{-1/2} \\
    &\qquad +  H_{2,W,\Lambda},
    \end{split}
\end{equation}

with:
\begin{align*}
    H_{2,W,\Lambda} &= W^{-1/2} X_{2,W} L_{2,W,\Lambda}W^{-1/2}\\
    &= W^{-1/2} X_{2,W}\biggl(\Lambda^{-1} - \Lambda^{-1}X_{2,W}^T(I_{n\times n} + P_{1,W} X_{2,W} \Lambda^{-1} X_{2,W}^T)^{-1} P_{1,W} X_{2,W} \Lambda^{-1}\biggr) \\
    &\qquad \cdot X^T_{W,2}P_{1,W} W^{-1/2}\\
    &= W^{-1/2} \Gamma_{W, \Lambda}P_{1,W}W^{-1/2} - W^{-1/2} \Gamma_{W, \Lambda}(I_{n\times n} + P_{1,W} \Gamma_{W, \Lambda})^{-1}P_{1,W}\Gamma_{W, \Lambda}P_{1,W}W^{-1/2}\\
    &= W^{-1/2} \Gamma_{W, \Lambda}\biggl(I_{n\times n} - (I_{n\times n} + P_{1,W} \Gamma_{W, \Lambda})^{-1}P_{1,W}\Gamma_{W, \Lambda}\biggr)P_{1,W} W^{-1/2},
\end{align*}
which completes the proof.

\subsection{Prediction for new samples and estimation of coefficients}
In line with the expression for $H_{W,\Lambda}$ (\eqref{Hmatunpen}) a prediction hat matrix $H^{\text{new}}_{W,\Lambda} =  X^{\text{new}} (X^T W X+\Lambda')^{-1}X^T$ is easily computed:
\begin{equation}\label{Hatpred}
H^{\text{new}}_{W,\Lambda} = X_1^{\text{new}} K_{W,\Lambda} + \Gamma_{\Lambda}^\text{new} M_{W,\Lambda},
\end{equation}
where
$$\Gamma_{\Lambda}^\text{new} = \sum_{b=1}^B \lambda^{-1}_b \Sigma_{2,b}^\text{new},\ \  \Sigma_{2,b}^\text{new} = X^{\text{new}}_{2,b} (X_{2,b})^T,$$

$$K_{W,\Lambda} = (X_{1,W}^TX_{1,W})^{-1}X_{1,W}^T \left( I_{n\times n} -  W^{1/2}H_{2,W,\Lambda}W^{1/2}\right)W^{-1/2}$$
and
$$M_{W,\Lambda} =  W^{1/2} \biggl(I_{n\times n} - (I_{n\times n} + P_{1,W} \Gamma_{W, \Lambda})^{-1}P_{1,W}\Gamma_{W, \Lambda}\biggr)P_{1,W} W^{-1/2}.$$
Here, $K_{W,\Lambda}$ and $M_{W,\Lambda}$ are $p_1 \times n$ and $n \times n$ matrices (with $p_1 < n$) available from the fitting, as part of $H_{W,\Lambda}$, which are
easily stored. In addition, the $n$ vector containing the final linearized response $\bm{L} = \bm{C} + W \boldeta$ (\eqref{eta}) of the IWLS algorithm needs to be stored to compute linear predictors for the new samples:
$\boldeta^{\text{new}}_{W,\Lambda} = H^{\text{new}}_{W,\Lambda} \bm{L}.$

\para
For estimation of $\bbeta$, we  use that $X\bbeta_{\Lambda} = H_{W,\Lambda}\bm{L}, $ and analogous to \eqref{Hatpred} we have $$H_{W,\Lambda} = X_1 K_{W,\Lambda} + X_2\Lambda^{-1} X_2^T M_{W,\Lambda} =
X  \left[\begin{array}{c}
    K_{W,\Lambda}  \\
    \Lambda^{-1} X_2^T M_{W,\Lambda}
    \end{array}
    \right].
$$
Therefore,
$$\hat{\bbeta}_{\Lambda} = \left[\begin{array}{c}
    K_{W,\Lambda}  \\
    \Lambda^{-1} X_2^T M_{W,\Lambda}
    \end{array}
    \right] \bm{L}.$$

\subsection{Cox ridge}
In Cox survival regression, the outcome $Y_i=(t_i,d_i), i=1,..,n$ denotes at which time $t_i$ an event occurred, $d_i=1$, or was censored, $d_i=0$. Details for fitting Cox ridge regression by Newton updating (and hence IWLS) are given in \cite[]{Houwelingen2006}. For the use of the IWLS algorithm, it suffices to replace the CVL ( \eqref{CVL}) by the cross-validated Cox likelihood \cite[]{Houwelingen2006} and update $W$ and $C$; all other formulas remain unchanged. Note that, as outlined by \cite{Meijer2013}, it is convenient to use the \emph{full} log-likelihood, and not the partial one, because the latter renders a non-diagonal weight matrix.
The penalized full log-likelihood is:
\begin{equation}\label{Coxpenlik}
\ell(\bbeta;\bY,X,\Lambda) = \sum_{i=1}^n\left(d_i (\log(h_0(t_i)) + \boldeta_i) - H_0(t_i)\exp(\boldeta_i)\right) - \frac{1}{2}\sum_{b=1}^B \lambda_b ||\bbeta_b||_2^2,
\end{equation}
with linear predictor $\boldeta_i = X_i\bbeta$. As for the GLM case, $\ell()$ is maximized by use of the IWLS algorithm, analogous to \eqref{tildeY} to \eqref{eta}.
In this case, updating of the baseline hazard is also required.
First, denote the hazard function for individual $i$ by $h_i(t)$, which by assumption is proportional to a baseline hazard $h_0(t)$ with cumulative hazard $H_0(t)$:
\begin{align}
    h_i(t)=h_0(t)\exp(\boldeta_i),\ \ H_0(t)=\int_{s=0}^t h_0(s)\, ds.
\end{align}

The IWLS algorithm for maximizing \eqref{Coxpenlik} then becomes:
\begin{alignat}{2}
\hat{H}_0(t) &=\sum_{i:\ t_i\leq t}\hat{h}_0(t_i),\ \hat{h}_0(t_i)=d_i\biggl(\sum_{j:\ t_j\geq t_i} \exp(\boldeta_j)\biggr)^{-1}  &\text{(Breslow estimator)}\\
W &= \text{diag}((w_i)_{i=1}^{n}), w_i =  H_0(t_i)\exp(\boldeta_i) &\text{ (sample weights)}\\
\bm{C} &= (c_i)_{i=1}^{n}, c_i = d_i - w_i,\ \  &\text{ (centered response vector)}\\
H_{\Lambda, W} &= X (\Lambda + X^T W X)^{-1} X^T  &\text{ (hat matrix)}\\
\boldeta^{\text{update}} &= \boldeta^{\text{update}}_{\Lambda, W} =  H_{\Lambda,W}(\bm{C} + W \boldeta) &\text{ (linear predictor update)}.
\end{alignat}
As before, this depends only on the linear predictor $\boldeta$, not on $\bbeta$.

\subsection{Functionality of the \texttt{multiridge} package}
The \texttt{multiridge} package has the following functionalities:
\begin{itemize}
\item Fit of multi-lambda ridge by IWLS algorithm for linear, logistic and Cox ridge regression
\item Inclusion of non-penalized covariates
\item Fast SVD-based CV per data type to initialize multi-lambda optimization
\item Optimization of $\Lambda$ using any of the following criteria, all cross-validated: log-likelihood (all), AUC (binary), mean-squared error (linear, binary), c-index (survival)

\item Optimization of $\Lambda$ using \texttt{mgcv}'s marginal likelihood approximation; only for settings with all covariates penalized

\item Prediction on new samples, and computation of final coefficients $\hat{\bbeta}_{\Lambda}$ for converged $\Lambda$
\item Two-stage preferential ridge
\item Paired ridge, which can be combined with multi-lambda ridge such that two data types are paired
\item Repeated cross-validation for penalty parameter tuning
\item Double (repeated) cross-validation for performance evaluation
\end{itemize}

Dependencies are:
\begin{itemize}
\item \texttt{penalized}: performing single-lambda CV after SVD (for initialization)
\item \texttt{mgcv}: marginal likelihood approximation
\item \texttt{pROC} and \texttt{risksetROC}: computing performance metrics AUC and c-index for binary and survival response, respectively.
\item \texttt{snowfall}: parallel computing
\end{itemize}

\subsection{Details about the data}
Below we give further details on the data used in this paper and in the \texttt{multiridge} package.

\subsection{TCGA data}
Overall survival, miRNA, mRNA and CNV data was retrieved from The Cancer Genome Atlas (\url{https://tcga-data.nci.nih.gov/tcga/}) using \texttt{TCGAbiolinks} \cite[]{colaprico2015tcgabiolinks}.
We considered all complete samples from the following tumor types:
\begin{itemize}
\item Mesothelioma \cite[MESO;][]{tcga2018meso}, $n=84$
\item Kidney renal clear cell carcinoma \cite[KIRC;][]{tcga2013kirc}, $n=506$
\item Sarcoma \cite[SARC;][]{tcga2017sarc}, $n=253$
\item Thymoma \cite[THYM;][]{tcga2018thym}, $n=118$
\end{itemize}
Data was preprocessed as described previously in \cite{rauschenberger2019sparse}.
Processed data are available via \url{https://github.com/markvdwiel/multiridge}.

\subsection{Cervical cancer data}
For the cervical cancer data, we retrieved molecular data corresponding to two class labels: controls and cases. Here, the cases are women with a last-stage precursor lesion for cervical cancer, a so-called cervical intraepithelial neoplasia, stage 3 (CIN3). Whereas lower-grade precursor lesions are known to regress back to normal, this higher grade has a relatively high risk to progress to cancer, and is therefore usually surgically removed. Hence, control versus CIN3 is the relevant classification problem. The molecular data were obtained from self-samples, as described in \cite{verlaat2018identification} for methylation and in \cite{snoek2019genome} for miRNA, which both include details on the data preprocessing. We matched the molecular samples from the same individuals, rendering $n=43$ samples (25 controls and 18 cases) for which both molecular data types were available.
Processed data are available as part of the demo via \url{https://github.com/markvdwiel/multiridge}.

\subsection{Coeficients \texttt{multiridge} vs \texttt{multiridge\_pref}}
\begin{table}[ht]
\centering
\begin{tabular}{r|rrr||rr|r}
  \hline
 & \multicolumn{3}{|c||}{\texttt{multiridge}} & \multicolumn{3}{c}{\texttt{multiridge\_pref}}\\
 & miRNA  & CNV & mRNA & miRNA  & CNV & mRNA\\
  \hline
MESO & 1.67  & 0.65 & 5.52 & 7.41  & 0.22 & 2.82 \\
  KIRC & 6.17  & 0.36 & 2.49 & 8.88  & 0.03 & 1.85\\
  SARC & 1.57  & 3.06 & 4.45 & 2.73  & 2.97 & 3.62\\
  THYM & 10.21  & 0.13 & 0.48 & 9.98 & 0.05  & 0.00 \\
   \hline
\end{tabular}\caption{Median (across 3 splits) sum of absolute coefficients ($||\bbeta||_1$) per data set and type}\label{absbetas}
\end{table}

%


\subsection{Marginal likelihood derivation}
Extending our earlier work \cite{veerman2020estimation} (single penalty case) to a penalty matrix $\Lambda$, we have for the marginal likelihood:
\begin{equation}\label{marglik}
\ML(\Lambda) =  \int_{\bbeta \in \mathbb{R}^p} \Lik(\bY; \bbeta,X) \pi_{\Lambda}(\bbeta) d\bbeta
\end{equation}
where $\pi_{\Lambda}(\bbeta)$ denotes the product prior the $j$th component of which is the central Gaussian distribution with variance $1/\lambda_j$. A simple, but crucial observation is that for GLM (and Cox models):
\begin{equation}\label{marglik2}
\ML(\Lambda) = E_{\pi_{\Lambda}(\bbeta)}[\Lik(\bY; \bbeta,X)] = E_{\pi_{\Lambda}(\bbeta)}[\Lik(\bY; X\bbeta, I_n)] =
E_{\pi'_{\Lambda}(\boldeta)}[\Lik(\bY; \boldeta, I_n)],
\end{equation}
because the likelihood depends on $\bbeta$ only via the linear predictor $\boldeta = X\bbeta$.
Here, $\pi'_{\Lambda}(X\bbeta)$ is the implied $n$-dimensional prior distribution of $\boldeta = X\bbeta$, which is a central multivariate normal with covariance matrix
$\Gamma_\Lambda= X\Lambda^{-1}X^T$.

\subsection{Multi-penalty Bayesian probit regression}
The reparametrization $\bbeta \rightarrow \boldeta$ to gain computational efficiency applies to other versions of multi-penalty ridge regression as well.
For the linear case, \cite{perrakis2019scalable} use equality \eqref{gamma} to derive very efficient algorithms for obtaining posteriors from Bayesian multi-penalty ridge regression.
Here, we extend this to Bayesian probit ridge regression for handling binary response by reformulating the variational Bayes (VB) algorithm in
\cite[]{ormerod2010explaining}.
An auxiliary variable is introduced to make VB steps tractable. The model with binary response $\bY\in\mathbb{R}^n$, auxiliary variable $\bs{a}\in\mathbb{R}^n$, observed data $X\in\mathbb{R}^{n\times p}$ and regression coefficients $\bs{\beta}\in\mathbb{R}^p$ is given by:
\begin{align*}
    &y_i|a_i \overset{ind.}{\sim} \pi(y_i|a_i) = I(a_i\geq 0)^y_iI(a_i< 0)^{1-y_i}\\
    &a_i|\bs{\beta} \overset{ind.}{\sim} N((X\bs{\beta})_i,1)\\
    &\bs{\beta}|\bs{\mu}_\beta,\Sigma_\beta \sim N(\bs{\mu}_\beta,\Sigma_\beta).
\end{align*}
We consider prior mean $\bs{\mu}_\beta=\bs{0}$ and $\Sigma_\beta=\Lambda^{-1}$, as in multi-penalty ridge regression.
The posterior $q$ is approximated under the assumption that the posterior factorizes over $\bs{a}$ and $\bs{\beta}$: $q(\bs{a},\bs{\beta})=q_a(\bs{a})q_\beta(\bs{\beta})$. The VB estimates $q^*_a,q^*_\beta$ are then analytical and are found by iteratively updating the posterior means of $\bs{a}$ and $\bs{\beta}$, denoted by $\bs{\mu}_{q(a)}$ and $\bs{\mu}_{q(\beta)}$ \cite[Algorithm 4]{ormerod2010explaining}.
As the update for $\bs{\mu}_{q(a)}$ only depends on $\bs{\mu}_{q(\beta)}$ via $\bs{\mu}_{q(\eta)} = X\bs{\mu}_{q(\beta)}$, the iterative steps in their Algorithm 4 can be conveniently rewritten in lower-dimensional computations. Recall that the hat-matrix $H_\Lambda= H_{\Lambda,I_n} = X(X^TX+\Lambda)^{-1}X^T$ is efficiently computed using \eqref{gamma} and \eqref{Hmateff}. Algorithm \ref{alg:VBprobit} then reformulates Algorithm 4 of \cite{ormerod2010explaining} to the $n$-dimensional (hence low-dimensional) space.
\begin{algorithm}
Initialize $\bs{\mu}_{q(a)}$.\\
Cycle:
\begin{algorithmic}
\STATE $\bs{\mu}_{q(\eta)} \gets X(X^TX+\Lambda)^{-1}X^T\bs{\mu}_{q(a)} = H_\Lambda \bs{\mu}_{q(a)}$
\STATE $\bs{\mu}_{q(a)} \leftarrow \bs{\mu}_{q(\eta)} + \dfrac{\phi(\bs{\mu}_{q(\eta)})}{\Phi(\bs{\mu}_{q(\eta)})^{\bY}(\Phi(\bs{\mu}_{q(\eta)})-\bs{1}_n)^{\bs{1}_n-\bY}},$
\end{algorithmic}
where $\Phi$ denotes the standard normal CDF.
\caption{VB scheme to approximate posteriors for Bayesian probit ridge regression in $n$-dimenisonal space}
\label{alg:VBprobit}
\end{algorithm}
Note the similarities of Algorithm \ref{alg:VBprobit} with the IWLS algorithm: an auxiliary variable is used to `linearize' the response ($W$ for IWLS), and alternating updating is required. VB algorithms use a lower bound of the marginal likelihood, the evidence lower-bound (elbo) to monitor convergence. In addition, analogously to CVL in the frequentist approach, the elbo may be used as a maximization criterion to estimate hyperparameters, here $\Lambda$, rendering a variational Bayes-empirical Bayes algorithm \cite[]{WielEB}.
\cite{ormerod2010explaining} present an expression for the elbo in terms of $\bbeta$. Some algebra, detailed below, shows that the elbo can also be entirely expressed in terms of (the current estimates of)
$\bs{\mu}_{q(\eta)}, \bs{\mu}_{q(a)}$, $H_{\Lambda}$  and $\Gamma_{\Lambda}$, which are efficiently  computed for multiple $\Lambda$'s using \eqref{gamma} and \eqref{Hmateff}. This facilitates fast optimization of the elbo in terms of $\Lambda$.
Once $\Lambda$ is fixed, Algorithm \ref{alg:VBprobit} produces predictions $\bs{\mu}_{q(\eta)}$ and their uncertainties: the approximate posterior of $\bs{\eta}=X\bs{\beta}$ is $N(\bs{\mu}_{q(\eta)},H_\Lambda)$, which is implied by the posterior of $\bs{\beta}$ \cite[]{ormerod2010explaining}. Predictions and uncertainities on the probabilistic scale, $\Phi(\bs{\eta})$, follow straightforwardly.

Finally, to use Algorithm \ref{alg:VBprobit} in practice a Bayesian evaluation of predictive performance is required. Given that the elbo is an approximate likelihood in terms of the hyperparameters, it can not be used to compare models with a different number of hyperparameters, e.g. corresponding to a different number of data types in our setting. Instead, we propose using the Bayesian counterpart of CVL, the conditional predictive ordinate (CPO). This is the geometric mean of $\pi(Y_i|\bY_{-i})$: the likelihood of $Y_i$ given (a model trained on) $\bY_{-i}$, where `training' includes hyperparameter optimization. As we show in the Supplementary Material, the CPO can also be expressed in terms of $\bs{\eta}$, and only requires computing $n$ one-dimensional integrals. This facilitates fast Bayesian performance evaluation.

\subsection{Expression for the elbo}
The evidence lower bound, denoted by $\underline{p}(\bY;q)$,  is given in \cite{ormerod2010explaining}, and can be rewritten in low-dimensional computations as well, using that $\bs{\mu}_{q(\beta)}=(X^TX+\Lambda)^{-1}X^T \bs{\mu}_{q(a)}$ after convergence of the algorithm:
\begin{equation}\label{elbo}
\begin{split}
    \underline{p}(\bY;q) &= \bY^T\log\left(\Phi(\bs{\mu}_{q(\eta)})\right) + (\bs{1}_n-\bY)^T\log\left(\bs{1}_n-\Phi(\bs{\mu}_{q(\eta)}) \right) \\
    &\qquad - \frac{1}{2} \bs{\mu}_{q(a)}^TX(X^TX+\Lambda)^{-1}\Lambda (X^TX+\Lambda)^{-1}X^T\bs{\mu}_{q(a)} \\
    &\qquad - \frac{1}{2}\log|\Lambda^{-1} X^TX+I_{p\times p}|\\
    &= \bY^T\log\left(\Phi(\bs{\mu}_{q(\eta)})\right) + (\bs{1}_n-\bY)^T\log\left(\bs{1}_n-\Phi(\bs{\mu}_{q(\eta)}) \right) \\
    &\qquad - \frac{1}{2} \bs{\mu}_{q(a)}^TH_\Lambda(I_{n\times n}- H_\Lambda)\bs{\mu}_{q(a)} - \frac{1}{2}\log|X\Lambda^{-1} X^T+I_{n\times n}|\\
    &= \bY^T\log\left(\Phi(\bs{\mu}_{q(\eta)})\right) + (\bs{1}_n-\bY)^T\log\left(\bs{1}_n-\Phi(\bs{\mu}_{q(\eta)}) \right) \\
    &\qquad - \frac{1}{2} \bs{\mu}_{q(a)}^TH_\Lambda(I_{n\times n}- H_\Lambda)\bs{\mu}_{q(a)} - \frac{1}{2}\log|\Gamma_\Lambda+I_{n\times n}|,
    \end{split}
\end{equation}
with $\Gamma_\Lambda$ as in \eqref{gamma}. Here we used Sylvester's determinant identity \cite{sylvesteridentitypress2005applied} and:
\begin{align*}
    &X(X^TX+\Lambda)^{-1}\Lambda (X^TX+\Lambda)^{-1}X^T \\
    &\qquad = X(X^TX+\Lambda)^{-1}(X^TX + \Lambda - X^TX) (X^TX+\Lambda)^{-1}X^T\\
    &\qquad = X(I-(X^TX+\Lambda)^{-1}X^TX)(X^TX+\Lambda)^{-1}X^T\\
    &\qquad = H_\Lambda - H_\Lambda H_\Lambda=H_\Lambda(I_{n\times n}-H_\Lambda).
\end{align*}

\subsection{Expression for CPO}
The conditional predictive ordinate on log-level equals
\begin{equation}\label{cpo}
\begin{split}
\text{CPO}_{\text{log}} &= \frac{1}{n}\sum_{i=1}^n \log(\text{CPO}_i), \text{ with:}\\
\text{CPO}_i^{-1} &= \frac{1}{\pi(Y_i|\bY_{-i})} = \frac{\pi(\bY_{-i})}{\pi(\bY)} = \int_{\eta_i} \frac{\pi(\bY_{-i} | \eta_i) \pi(\eta_i)}{\pi(\bY)}d\eta_i\\
&= \int_{\eta_i} \frac{\pi(\bY | \eta_i)/\pi(Y_i | \eta_i) \pi(\eta_i)}{\pi(\bY)}d\eta_i=\int_{\eta_i} \frac{\pi(\eta_i|\bY) }{\pi(Y_i | \eta_i)}d\eta_i.
\end{split}
\end{equation}
Here, the posterior $\pi(\eta_i|\bY)$, with $\eta_i = \bX_i\bbeta^{\hat{\Lambda}_{(-i)}}_{-i}$, follows directly from the Gaussian posterior of $\bbeta^{\hat{\Lambda}_{(-i)}}_{-i}$ \cite[]{ormerod2010explaining}, where the empirical Bayes estimate of $\Lambda$, obtained by maximizing the elbo  $\underline{p}(\bY;q)$ \eqref{elbo}, and the posterior estimate of $\bbeta$ are obtained without use of sample $i$.
So, computation of CPO only requires one-dimensional numerical integration on top of the fast VB-EB method which combines fast approximation of $\pi(\eta_i|\bY)$ by Algorithm \ref{alg:VBprobit} (Main Document) with efficient computation (and maximization) of the elbo using \eqref{elbo}. As defined in \eqref{cpo} $\text{CPO}_{\text{log}}$ mimics leave-one-out-cross-validation, but this may straightforwardly be extended to $k$-fold CV, which will imply an additional speed-up of the algorithm, as fewer fits are required.
Finally, posterior $\pi(\eta_i|\bY)$  depends on the cross-validated versions of $H_{\Lambda}$ and $\Gamma_\Lambda$, which are efficiently computed from \eqref{Hmatout} and \eqref{gammasub}.

\newpage
\subsection{Additional Figures}
\subsection{Profile plot for CVL and AUC}
\begin{figure}
\includegraphics[scale=0.45]{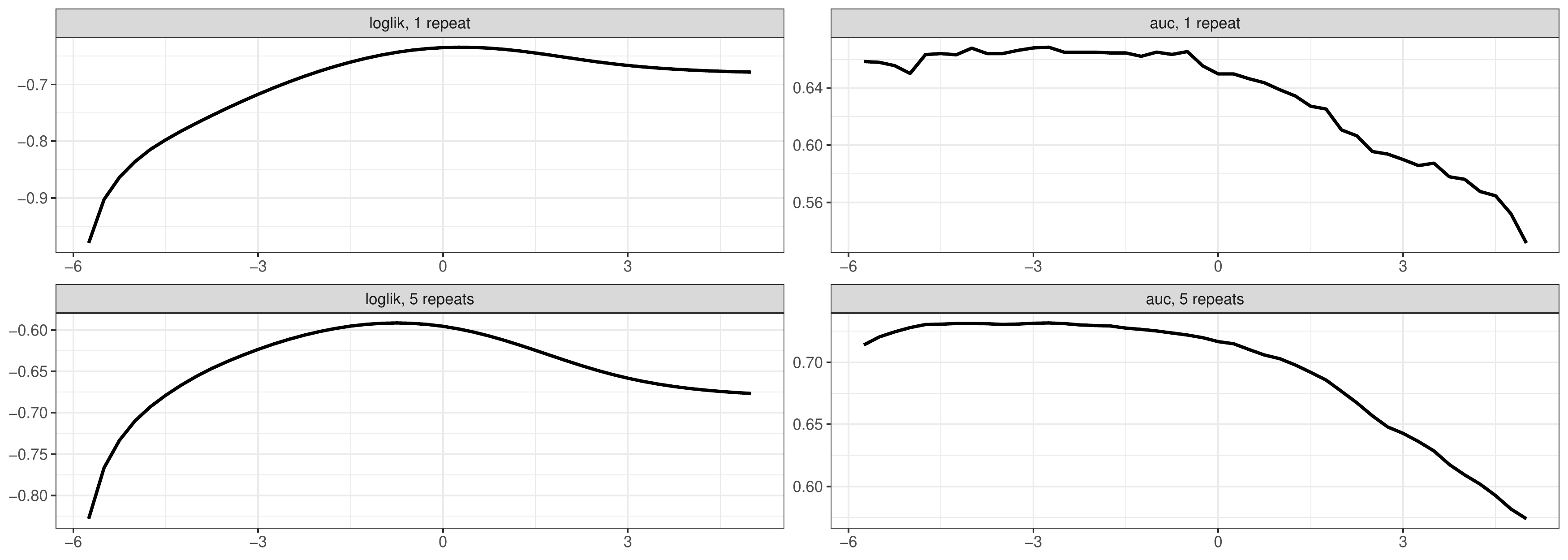}
\caption{Profile plots showing relative penalty parameter (log-scale; x-axis) versus cross-validated performance (log-likelihood, AUC; y-axis), using either one or five repeats of 5-fold CV. Penalty parameter relative to the one optimized for log-likelihood using 1 repeat.}\label{fig:profile}
\end{figure}

We used logistic ridge for methylation data presented in \cite{verlaat2018identification} ($n=43$, $p=365,620$) to illustrate profile plots
of cross-validated log-likelihood and AUC as a function of $\lambda$. Figure \ref{fig:profile} shows that the latter is not smooth when only one repeat of CV is used, but this can be countered by increasing the number of CV repeats.

\subsection{Comparison \texttt{multiridge\_ML} with \texttt{multiridge\_CV}}

\begin{figure}
\includegraphics[scale=0.45]{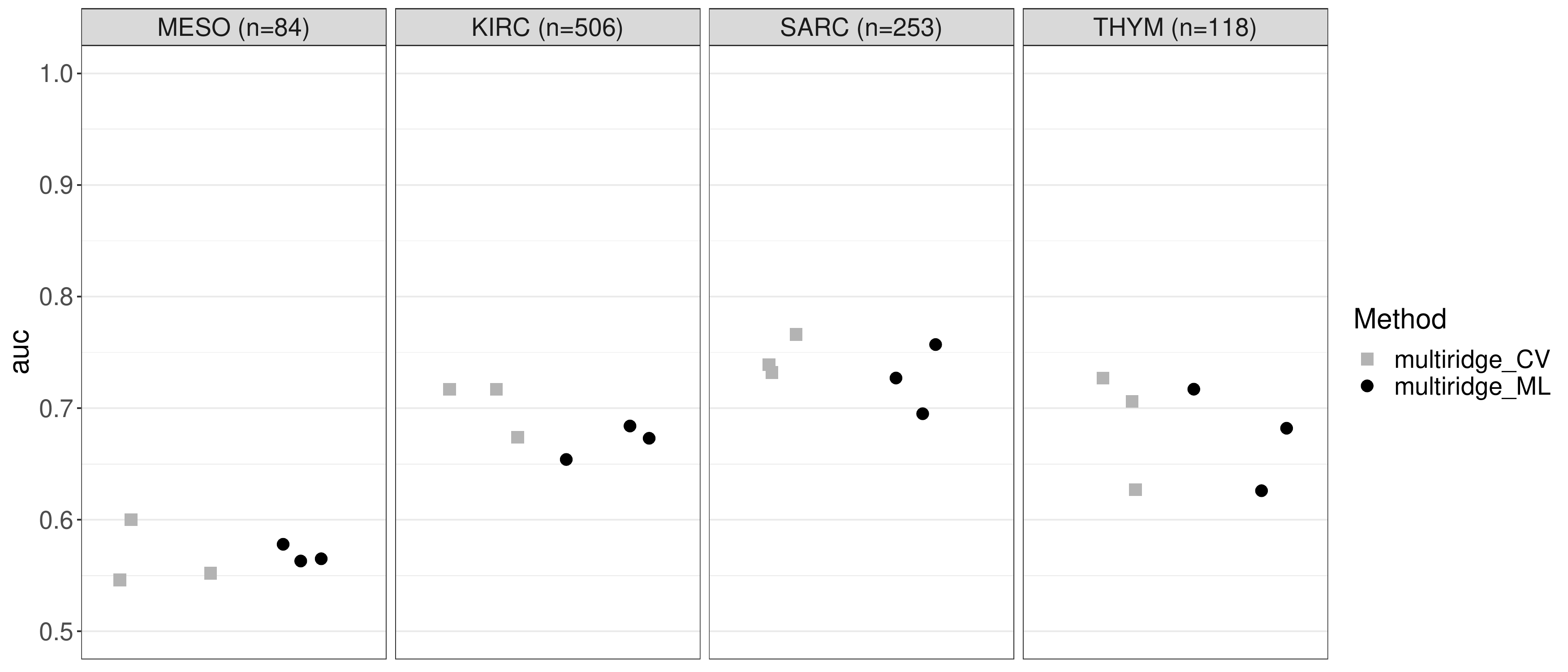}
\caption{Predictive performance for three splits per data set, as measured by AUC (y-axis). First setting: $\blambda = (20,1000,10000)$}\label{fig:perfmgcv_bin1}
\end{figure}

\begin{figure}
\includegraphics[scale=0.45]{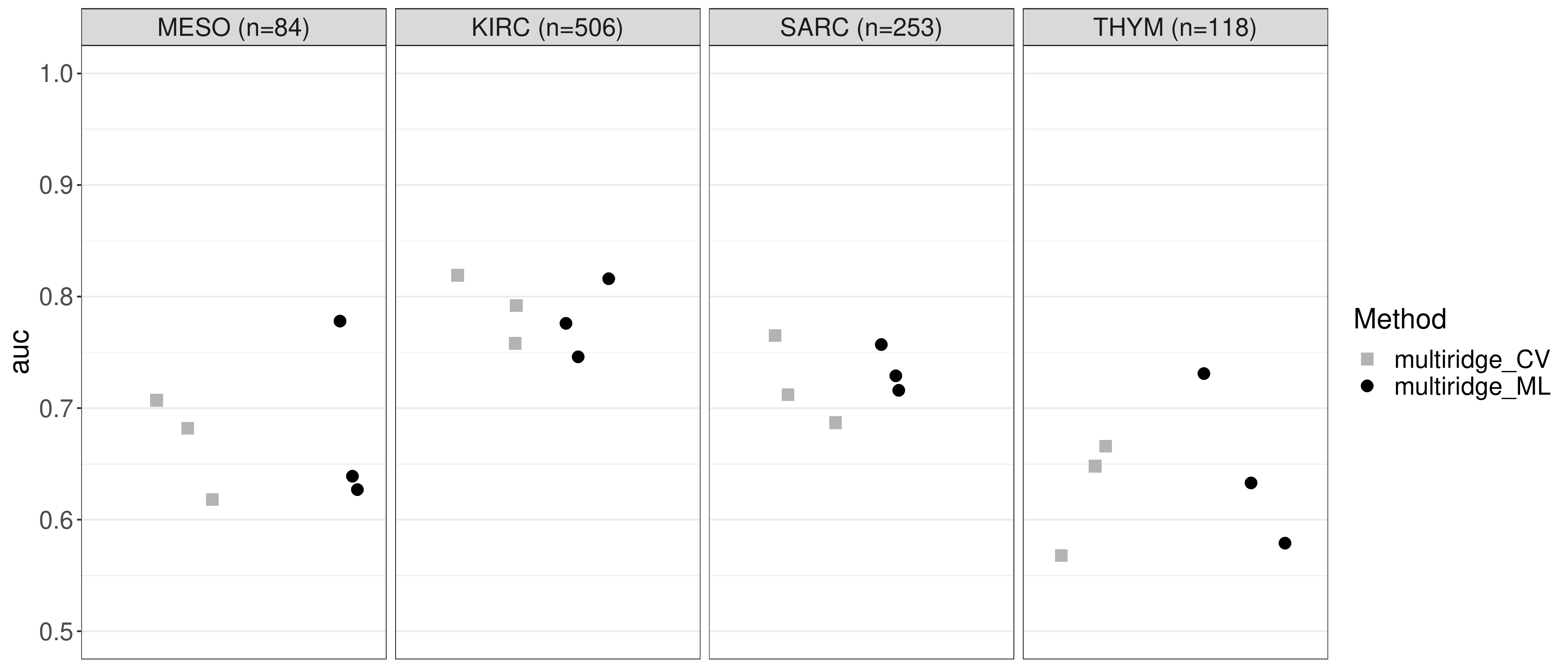}
\caption{Predictive performance for three splits per data set, as measured by AUC (y-axis). Second setting: $\blambda = c(100,200,2000)$}\label{fig:perfmgcv_bin2}
\end{figure}

\begin{figure}
\begin{center}
\includegraphics[scale=0.5]{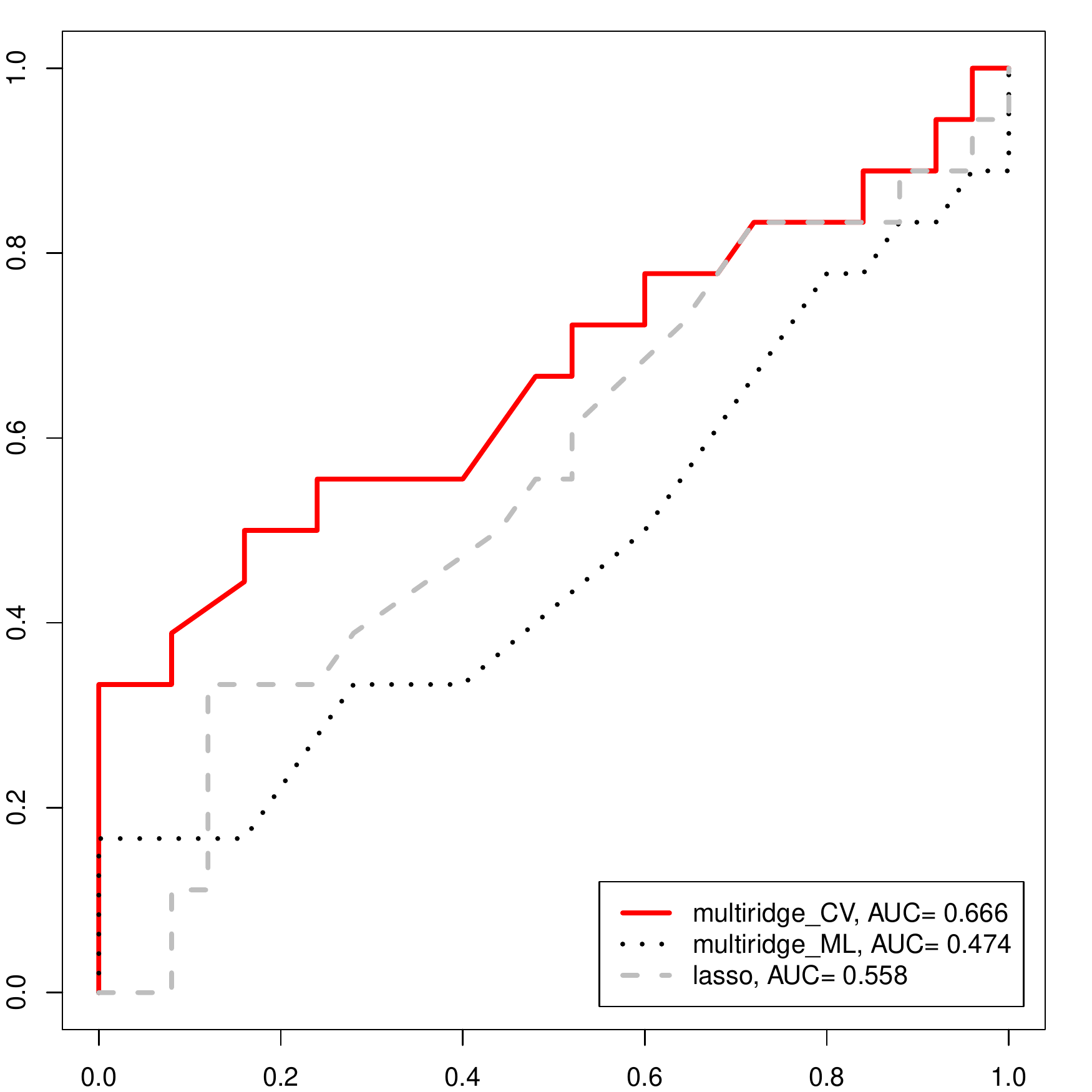}
\caption{ROC curves for classification of control vs cervical cancer precursor (CIN3)}\label{fig:roccervical}
\end{center}
\end{figure}

Figures \ref{fig:perfmgcv_bin1} and \ref{fig:perfmgcv_bin2} compare the predictive performances of \texttt{multiridge\_ML}  with that of \texttt{multiridge\_CV} for four TCGA multi-omics data set with simulated binary response. Figures \ref{fig:roccervical} compares the classification performances of \texttt{multiridge\_CV}, \texttt{multiridge\_ML} and \texttt{lasso} in the logistic regression setting.
Data consists of methylation and miRNA profiles for 43 samples, (25 controls and 18 cases).

\bibliographystyle{agsm}

\bibliography{C://Synchr//Bibfiles//bibarrays}      

\end{document}